\documentclass[aps,pra,twocolumn,showpacs,superscriptaddress,floatfix]{revtex4-1}
\usepackage{graphicx,amsmath,amssymb}
\usepackage[usenames]{color}
\usepackage[dvipsnames]{xcolor}
\usepackage[unicode=true,pdfusetitle,
 bookmarks=false,
 breaklinks=false,pdfborder={0 0 1},backref=false,colorlinks,urlcolor=Black, linkcolor=blue,citecolor=blue]
 {hyperref}
\begin{document}

\title{Upgrading Quantum Metrology by Combined Sensitivity Resources in Mixed Linear-Nonlinear Light-Matter Interactions with Bias Field
}
\author{Zu-Jian Ying}
\email{yingzj@lzu.edu.cn}
%\affiliation{School of Physical Science and Technology, Key Laboratory for Quantum Theory and Applications of MoE,
%and Lanzhou Center for Theoretical Physics, Lanzhou University, Lanzhou 730000, China}
\affiliation{School of Physical Science and Technology, Lanzhou University, Lanzhou 730000, China}
\affiliation{Key Laboratory for Quantum Theory and Applications of MoE, Lanzhou Center for Theoretical Physics, Lanzhou University, Lanzhou 730000, China}

\begin{abstract}
The major goal of quantum metrology (QM) is to exploit the quantum resources to raise the measurement precision (MP) as high as possible. When the quantum resources such as squeezing has been widely explored, light-mater interaction systems set up a highly controllable platform applicable for QM in novel pursuit of high MP. However, critical QM by the conventional linear interaction is confronted with the restriction of low-frequency-limit condition and the detrimental problem of diverging preparation time of the probe state (PTPS). This work shows that mixed interactions by linear and nonlinear light-matter couplings in the presence of bias field can provide various quantum resources, including squeezing, degeneracy lifting, displacement and quantum phase transition. These resources manifest high sensitivity for QM as demonstrated by analytically obtained critical components or exponential behavior of quantum Fisher information. We find that these sensitivity resources can be combined to upgrade the upper bound of MP by many orders over the widely-applied squeezing resource. As further advantages, such an upgraded metrology protocol not only breaks the frequency-limit restrictions but also avoids the detrimental problem of diverging PTPS which were both encountered in linear interaction. Our work paves a way to exploit and combine all the resources in momentum, position and spin spaces to maximize the MP and expand the applicable conditions simultaneously.
\end{abstract}
\pacs{ }
\maketitle

%\date{\today}

\section{Introduction}

Recently increasing efforts have been exerted on the development of quantum
metrology (QM) which is a cornerstone of quantum
technologies~\cite{Garbe2020,Montenegro2021-Metrology,Chu2021-Metrology,Garbe2021-Metrology,Ilias2022-Metrology,
Ying2022-Metrology,Hotter2024-Metrology,Alushi2024PRL,Mukhopadhyay2024PRL,Mihailescuy2024,Ying-Topo-JC-nonHermitian-Fisher}.
On the other hand, with high controllability and tunability light-matter
interactions~\cite{Diaz2019RevModPhy,Kockum2019NRP,Eckle-Book-Models,JC-Larson2021,Boite2020,Qin-ExpLightMatter-2018,LiPengBo-Magnon-PRL-2024} are
building an ideal platform with great potential for applications in
QM~\cite{Garbe2020,Garbe2021-Metrology,Ilias2022-Metrology,Ying2022-Metrology,Hotter2024-Metrology,Ying-Topo-JC-nonHermitian-Fisher}.

The main quantum resources widely applied for QM~\cite%
{Degen2017-QuantSensing} include entanglement~\cite{Pezze2018entangelment}
and squeezing~\cite{Maccone2020Squeezing,Lawrie2019Squeezing,Gietka2023PRL-Squeezing,Gietka2023PRL2-Squeezing,Candeloro2021-Squeezing}.
Entanglement as a QM resource suffers from the difficulties of producing
and maintaining due to its vulnerability~\cite{Horodecki2009entanglement}
and also its detection is often complex and challenging~\cite%
{Guhne2009EntanglementDetection}. Squeezing, with its robustness to
decoherence and dissipation, has been mostly explored in QM~\cite%
{Maccone2020Squeezing,Lawrie2019Squeezing,Gietka2023PRL-Squeezing,Gietka2023PRL2-Squeezing,Candeloro2021-Squeezing}.
Since essentially the major aim of QM is to exploit the quantum resources
to realize the measurement precision (MP) as high as possible, it is still
desirable to explore novel quantum resources or approaches to attain higher
measurement precision. In this regards, critical QM is arising to open a
promising avenue for QM by exploiting the sensitivity in the critical
behavior of quantum phase transition (QPT)~\cite{Garbe2020,Garbe2021-Metrology,Ilias2022-Metrology,Ying2022-Metrology,Hotter2024-Metrology}.
It is worthy to note that QPT is not an exclusive phenomenon of
thermodynamical systems in condensed matter, actually in recent years it has
been found that finite-component systems in light-matter interactions can
also manifest QPTs~\cite{Liu2021AQT,Ashhab2013,Ying2015,Hwang2015PRL,Ying2020-nonlinear-bias,Ying-2021-AQT,LiuM2017PRL,Hwang2016PRL,Irish2017, Ying-gapped-top,Ying-Stark-top,Ying-Spin-Winding,Ying-2018-arxiv,Ying-JC-winding,Ying-Topo-JC-nonHermitian,Ying-Topo-JC-nonHermitian-Fisher,Ying-gC-by-QFI-2024, Ying-g2hz-QFI-2024, Grimaudo2022q2QPT,Grimaudo2023-Entropy,Grimaudo2024PRR,Zhu2024PRL,DeepStrong-JC-Huang-2024}.
Indeed, the fundamental model in lighter-matter interactions--the quantum
Rabi model~\cite{rabi1936,Braak2011,Rabi-Braak,Eckle-Book-Models}--possesses a QPT with the low-frequency limit as a replacement of the
thermodynamical limit in condense matter~\cite{Ashhab2013,Ying2015,Hwang2015PRL,LiuM2017PRL,Irish2017}. In fact, these two limit can be even bridged via critical scaling relation~\cite{LiuM2017PRL}. As the coupling in the quantum Rabi model is linear, the critical QM based on the quantum
Rabi model can be referred to as linear critical QM here.

Although the finite-component systems are not bothered by the difficulty to
reach a strict thermodynamical equilibrium required by the high precision in
QM of many-body systems, linear
critical QM is however
confronted with some other problems. These problems include limitation of the transition
to a single parameter point, requirement of the low-frequency
limit, and diverging preparation time of probe state (PTPT)~\cite{Garbe2020,Ying2022-Metrology,Gietka2022-ProbeTime},
which might hinder the applications of linear critical QM. Improved protocols of critical QM
surmounting these problems are in need.

Nevertheless, in the emerging phenomenology of light-matter interactions~\cite{Diaz2019RevModPhy,Kockum2019NRP,Eckle-Book-Models,JC-Larson2021,Boite2020,Qin-ExpLightMatter-2018,LiPengBo-Magnon-PRL-2024,Kockum2017,
Liu2021AQT,Ashhab2013,Ying2015,Hwang2015PRL,Ying2020-nonlinear-bias,Ying-2021-AQT,LiuM2017PRL,Hwang2016PRL,Irish2017,
rabi1936,Braak2011,Rabi-Braak,Solano2011,ChenQH2012,
Gietka2022-ProbeTime,
Ying-gapped-top,Ying-Stark-top,Ying-Spin-Winding,Ying-2018-arxiv,Ying-JC-winding,Ying-Topo-JC-nonHermitian,Ying-Topo-JC-nonHermitian-Fisher,Ying-gC-by-QFI-2024, Ying-g2hz-QFI-2024, Grimaudo2022q2QPT,Grimaudo2023-Entropy,Grimaudo2024PRR,Zhu2024PRL,DeepStrong-JC-Huang-2024,
Ulstrong-JC-2,Ulstrong-JC-3-Adam-2019,LiuGang2023,PRX-Xie-Anistropy,Irish-class-quan-corresp,Li2020conical,
PengJie2019,PengJ2021PRL,Yimin2018,Zheng2017,Yan2023-AQT,Padilla2022,Ma2020Nonlinear,AiQ2023,KuangLM2024AQT,
Gao2022Rabi-dimer,Gao2022Rabi-aniso,Chen-2021-NC,
Alushi2023PRX} finite-component QPTs can exhibit abundant exotic quantum
phenomena such as criticality and universality~\cite{Ashhab2013,Ying2015,Hwang2015PRL,LiuM2017PRL,Hwang2016PRL,Irish2017,Ying-2021-AQT,Ying-Stark-top},
multi-criticality~\cite{Ying2020-nonlinear-bias,Ying-2021-AQT,Ying-gapped-top,Ying-Stark-top,Ying-2018-arxiv, Ying-JC-winding,Zhu2024PRL,Lyu24-Multicritical,Wu24-RabiTransition-Exp}, compromise of universality and diversity~\cite{Ying-2021-AQT,Ying-Stark-top},
topological phase transitions conventionally with~\cite{Ying-2021-AQT,Ying-gapped-top,Ying-Stark-top} and unconventionally without~\cite{Ying-gapped-top,Ying-Stark-top,Ying-Spin-Winding,Ying-JC-winding}
gap closing, anti-level-crossing and spin knot states~\cite{Ying-Spin-Winding}, coexistence of
Landau-class and topological-class transitions~\cite{Ying-2021-AQT,Ying-Stark-top,Ying-JC-winding,Ying-Topo-JC-nonHermitian-Fisher},
robust topological feature against non-Hermiticity~\cite{Ying-Topo-JC-nonHermitian} and
universal criticality of exceptional points~\cite{Ying-Topo-JC-nonHermitian-Fisher}.
In particular, the nonlinear
interaction manifests various patterns of symmetry breaking which lead to
different types of QPTs~\cite{Ying2020-nonlinear-bias,Ying-2018-arxiv}.
These exotic properties might provide more potential quantum resources for
QM.

In the present work, we propose to explore and combine various sensitivity
resources to upgrade the upper bound of measurement precision in QM. Such a
protocol can be realized in a light-matter-interaction system with mixed
linear and nonlinear couplings in the presence of bias field. Indeed, from
the system we extract several quantum resources, including squeezing,
degeneracy lifting, displacement and QPT, all with high sensitivity. We
analytically obtain the critical components or exponential behavior of QFI
which represents the upper bound of measurement precision. We combine and
compare these resources step by step, showing that each combination can give
a dramatic boost to the enhancement of the QFI. Finally a broadest
combination of squeezing, displacement and QPT yields a maximized QFI with
an improvement by many orders over the widely-used squeezing resource.
Besides the upgrading of the upper bound of measurement precision, our protocol
of QM also shows several other advantages including more global parameter
regime, breaking the frequency-limit restriction, and avoiding the
detrimental diverging PTPS which were all encountered as problems in
linear critical QM. By a more general view in the Wigner
function, our work paves a way to exploit and combine all the resources in
momentum, position and spin spaces to upgrade the measurement precision and
simultaneously expand the applicable regime and conditions.

The paper is organized as follows. Section~\ref{Section-model} introduces and reformulates
the light-matter interaction model with mixed linear and nonlinear couplings
in the presence of bias field.
Section~\ref{Section-QFI}
defines and simplifies the QFI which characterizes the upper bound of measurement precision in QM.
Section~\ref{Section-wave-patterns}
describes different variation patterns of the wave function which provide various sensitivity resources for QM.
Section~\ref{Section-QM}
is devoted to combinations of different sensitivity resources to upgrade the measurement precision in QM, eventually with
enhancement of QFI by many orders.
Section~\ref{Section-advantages}
shows several more advantages of our QM protocol in extending parameter
application regime, breaking the frequency-limit restriction, and avoiding divergent PTPS.
Section~\ref{Section-Wigner}
analyzes the resources by the Wigner function to gain a more general view from the position,
momentum and spin spaces.
Finally, Section~\ref{Section-Conclusion} gives a summary of conclusions.

\section{Asymmetric linear-nonlinear-mixed Quantum Rabi Model}
\label{Section-model}

We consider a general nonlinear quantum Rabi model for light-matter
interactions~\cite{Ying2020-nonlinear-bias}
\begin{equation}
H=\omega a^{\dagger }a+\frac{\Omega }{2}\hat{\sigma}_{x}+g_{1}\hat{\sigma}
_{z}(a^{\dagger }+a)+g_{2}\hat{\sigma}_{z}(a^{\dagger }+a)^{2}-\epsilon \hat{
\sigma}_{z}.  \label{H-g1-g2-bias}
\end{equation}
which describes couplings between a quantized bosonic mode with frequency $\omega $,
created (annihilated) by $a^{\dagger }$ ($a)$, and a qubit
represented by the Pauli matrices $\hat{\sigma}_{x,y,z}$. Here $\Omega $
denotes the energy splitting of atom in cavity system and the tunneling or
spin flipping energy of flux qubit in superconducting circuit system~\cite{flux-qubit-Mooij-1999}. The
interactions include the conventional linear coupling with strength $g_{1}$~\cite{rabi1936,Braak2011,Rabi-Braak,Eckle-Book-Models}
and a nonlinear coupling with strength $g_{2}$~\cite{Felicetti2018-mixed-TPP-SPP,Felicetti2015-TwoPhotonProcess,e-collpase-Garbe-2017,Rico2020,e-collpase-Duan-2016,CongLei2019,Bertet-Nonlinear-Experim-Model-2005}.
The physical regime of $g_{2}$ lies in $[0,g_{\mathrm{T}}]$, while beyond
the spectral collapse point $g_{\mathrm{T}}$ the system is
unstable without lower energy bound~\cite{Felicetti2018-mixed-TPP-SPP,Felicetti2015-TwoPhotonProcess,e-collpase-Garbe-2017,Rico2020,e-collpase-Duan-2016,CongLei2019}.
Here the spectral collapse point of $g_{2}$ has shifted from $g_{\mathrm{t}}=\omega /2$ in the two-photon coupling $\hat{\sigma}_{z}[(a^{\dagger })^{2}+a^{2}]$ to $g_{\mathrm{T}}=\omega /4$ in the full quadratic coupling $\hat{\sigma}_{z}(a^{\dagger }+a)^{2}$~\cite{Ying-2018-arxiv,Ying2020-nonlinear-bias}.
Finally in $H$, the $\epsilon $ term is the bias field which can be tuned by external flux in superconducting circuit systems.

The linear quantum Rabi model has the symmetry $P_{1}=\hat{\sigma} _{x}e^{i\pi a^{\dagger }a}$.
Differently the nonlinear coupling in the two-photon form %$\hat{\sigma}_{z}[(a^{\dagger })^{2}+a^{2}]$
possesses the symmetry $P_{2}=\hat{\sigma} _{x}e^{i\pi a^{\dagger }a/2}$ which is however broken and replaced by
$P_{x}=e^{i\pi a^{\dagger }a}$ in the full quadratic form here. %$\hat{\sigma}_{z}(a^{\dagger }+a)^{2}$.
The bias is also an asymmetric term but there is a hidden symmetry at
certain points in linear-coupling situation~\cite{HiddenSymMangazeev2021,HiddenSymLi2021,HiddenSymBustos2021}. The coupling
mixing breaks all the symmetries and leads to rich QPTs~\cite{Ying2020-nonlinear-bias}.

By transformation
$a^{\dagger }=(\hat{x}-i\hat{p})/\sqrt{2}$,
$a=(\hat{x}+i \hat{p})/\sqrt{2}$ with position $x$ and momentum $\hat{p}=-i\frac{\partial}{\partial x}$,
we can rewrite $H$ as~\cite{Irish2014,Ying2015,Ying-2018-arxiv,Ying2020-nonlinear-bias}
$H_{x}=\sum_{\sigma _{z}=\pm }h_{\sigma _{z}}\left\vert \sigma
_{z}\right\rangle \left\langle \sigma _{z}\right\vert +\frac{\Omega }{2}
\sum_{\sigma _{z}=\pm }\left\vert \sigma _{z}\right\rangle \left\langle
\overline{\sigma }_{z}\right\vert $ where $\sigma _{z}=-\overline{\sigma }
_{z}=\pm $ labels the spin in $z$ direction. Here $h_{\pm }=\frac{\omega }{2}
\hat{p}^{2}+v_{\pm }\left( x\right) $ is the effective
singe-particle Hamiltonian, in the spin-dependent harmonic potential
\begin{equation}
v_{\pm }(x)=\frac{\omega }{2}\varpi _{\pm }^{2}\left( x\pm b_{\pm }\right)
^{2}+d_{\pm }(x)\mp \epsilon -\frac{\omega }{2},  \label{v-potentials}
\end{equation}
with frequency renormalization $\varpi _{\pm }$, displacement of potential
bottom $b_{\pm }$, coupling--mixing-induced asymmetry $d_{\pm }$,
\begin{eqnarray}
\varpi _{\pm } &=&\sqrt{1\pm g_{2}/g_{\mathrm{T}}}, \\
b_{\pm } &=&\frac{g_{1}^{\prime }}{1\pm g_{2}/g_{\mathrm{T}}}, \label{b-up-down}\\
d_{\pm } &=&-\frac{\overline{g}_{1}^{2}\Omega }{4(1\pm \overline{g}_{2})},
\label{v-d-g1}
\end{eqnarray}
and single-particle energy%
\begin{equation}
\varepsilon _{\pm }^{0}=\omega (n+\frac{1}{2})\sqrt{1\pm \overline{g}_{2}}
\mp \epsilon -\frac{1}{2}\omega .  \label{e0-up-down}
\end{equation}
In such a formalism the $\Omega $ term plays the role of spin flipping in
the spin space or tunneling in the position space~\cite%
{Irish2014,Ying2015,flux-qubit-Mooij-1999}. Hereafter we consider the ground
state which involves $n=0$.

%%%%%%%%%%%%%%%%%%%%%%%%%%%%%%%%%%%%%%%%%%%%%%%%%%%%%%%%%%%%%%%%%%%%%%%%%%%%%%%%%%%%%%%%%%%%%%%%%%
\begin{figure*}[t]
\includegraphics[width=2\columnwidth]{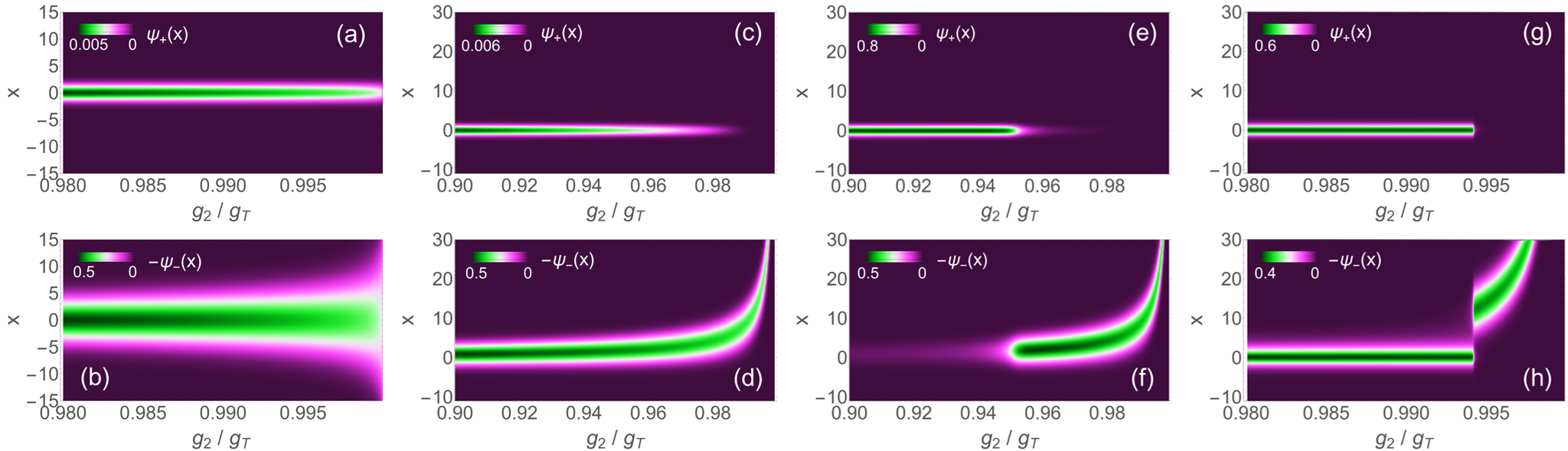}
\caption{Variation patterns of wave function and different sensitivity resources of quantum metrology.
Wave function components $\psi _{+}$ (a,c,e,g) and $\psi _{-}$ (b,d,f,h) with different parameters:
(a) $\epsilon =0$, $g_1 =0$, $\Omega =0.01\omega$, with squeezing resource.
(b) $\epsilon =0$, $g_1 =1.2\omega$, $\Omega =0.01\omega$, with squeezing and displacement resources.
(c) $\epsilon =0.33\omega$, $g_1 =1.2\omega$, $\Omega =0.01\omega$, with squeezing, displacement and transition resources.
(d) $\epsilon =0.33\omega$, $g_1 =0.1\omega$, $\Omega =1.0\omega$, with squeezing, displacement and transition resources at finite $\Omega$.
Here $\omega=1$ is set for the unit. There is a degeneracy-lifting pattern around $g_2 =0$ at $\epsilon =0$ and $g_1 =0$, with $\psi _{\pm}$ varying similar to (e) and (f) after the transition point except the absence of displacement.}
\label{fig-Wave-Function}
\end{figure*}
%%%%%%%%%%%%%%%%%%%%%%%%%%%%%%%%%%%%%%%%%%%%%%%%%%%%%%%%%%%%%%%%%%%%%%%%%%%%%%%%%%%%%%%%%%%%%%%%%%
%%%%%%%%%%%%%%%%%%%%%%%%%%%%%%%%%%%%%%%%%%%%%%%%%%%%%%%%%%%%%%%%%%%%%%%%%%%%%%%%%%%%%%%%%%%%%%%%%%
\begin{figure*}[t]
\includegraphics[width=2\columnwidth]{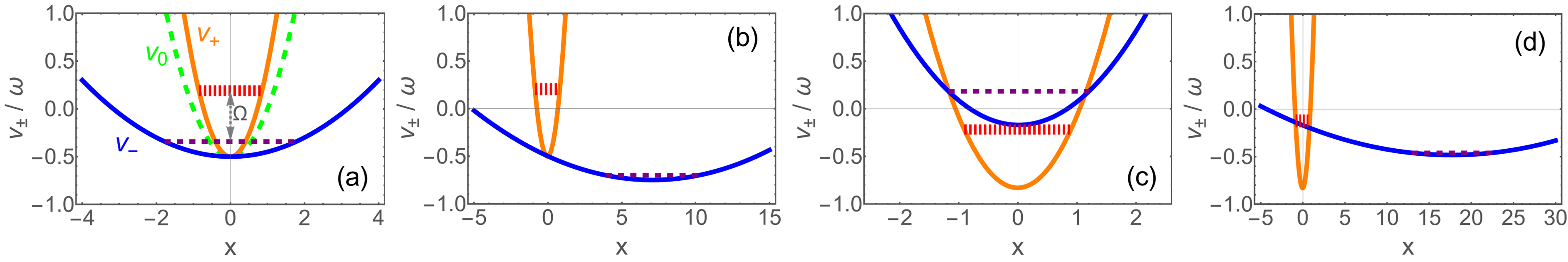}
\caption{Effective potential $v_\pm$ and physical picture for the squeezing (a), displacement (b) and transition (c,d).
(a) Narrowing of $v_{+}$ (orange solid) and broadening of $v_{-}$ (blue solid) compared with the original potential $v_0$ (green long-dashed) at $g_2=0.9g_{\rm T}$ with $g_1=0$ and $\epsilon=0$. The horizontal red dotted line and purple dashed line represent $\varepsilon_{+}$ and $\varepsilon_{-}$, while the arrows indicate the spin flipping $\Omega$.
(b) Displaced $v_{-}$ at $g_1=1.0g_{\rm s}$ with $g_2=0.9g_{\rm T}$ and $\epsilon=0$.
(c) Energy situation $\varepsilon_{-}>\varepsilon_{+}$ before the transition, at a finite bias $\epsilon=0.33\omega$ with $g_1=0.1g_{\rm s}$ and $g_2=0.5g_{\rm T}$.
(d) Reversed energy situation $\varepsilon_{-}<\varepsilon_{+}$ after the transition, at $g_2=0.998g_{\rm T}$ with $\epsilon=0.33\omega$ and $g_1=0.1g_{\rm s}$.  Here, $\Omega=0.01\omega$ in all panels and we set $\omega =1$ as the unit.
}
\label{fig-v}
\end{figure*}
%%%%%%%%%%%%%%%%%%%%%%%%%%%%%%%%%%%%%%%%%%%%%%%%%%%%%%%%%%%%%%%%%%%%%%%%%%%%%%%%%%%%%%%%%%%%%%%%%%

\section{Quantum Fisher Information (QFI) for quantum metrology}
\label{Section-QFI}

In QM the measurement precision of experimental estimation on a parameter $\lambda $ is bounded by $F_{Q}^{1/2}$~\cite{Cramer-Rao-bound},
where $F_{Q}$ is the QFI defined as \cite{Cramer-Rao-bound,Taddei2013FisherInfo,RamsPRX2018}
\begin{equation}
F_{Q}\left( \lambda \right) =4\left[ \langle \psi ^{\prime }\left( \lambda
\right) |\psi ^{\prime }\left( \lambda \right) \rangle -\left\vert \langle
\psi ^{\prime }\left( \lambda \right) |\psi \left( \lambda \right) \rangle
\right\vert ^{2}\right]   \label{Fq}
\end{equation}%
for a pure states $|\psi (\lambda )\rangle $. Here $^{\prime }$ denotes the
derivative with respect to the parameter $\lambda $. A higher QFI would mean
a higher measurement precision. For a real wave function $\psi (\lambda )$,
as is usually the case in non-degenerate states of a real Hamiltonian, the
QFI can be simplified to be~\cite{Ying-gC-by-QFI-2024}
\begin{equation}
F_{Q}=4\langle \psi ^{\prime }\left( \lambda \right) |\psi ^{\prime }\left(
\lambda \right) \rangle ,
\end{equation}%
which also applies for the ground state of our Hamiltonian (\ref{H-g1-g2-bias}) considered in the present work.

It is worth mentioning that the appearance of the QFI peak not only can be
employed for critical QM~\cite{Garbe2020,Garbe2021-Metrology,Ilias2022-Metrology,Ying2022-Metrology} but also
signals a QFT in fidelity theory~\cite{Zhou-FidelityQPT-2008,Zanardi-FidelityQPT-2006,Gu-FidelityQPT-2010,You-FidelityQPT-2007,You-FidelityQPT-2015},
as applied to identify the frequency dependence of the QPT in the quantum
Rabi model~\cite{Ying-gC-by-QFI-2024}. In fact, in an infinitesimal parameter
variation $\delta \lambda $ the fidelity $F$ can be expand as
\begin{equation}
F=\left\vert \langle \psi \left( \lambda \right) |\psi \left( \lambda
+\delta \lambda \right) \rangle \right\vert =1-\frac{\delta \lambda ^{2}}{2}
\chi _{F},
\end{equation}
thus the QFI is also the susceptibility of the fidelity by the
correspondence $\chi _{F}=F_{Q}/4$~\cite{Gu-FidelityQPT-2010,You-FidelityQPT-2007,You-FidelityQPT-2015}. In our
resources-combined QM addressed in the present work we will see peaks of QFI
which are really coming from the transition resource.

\section{Different variation patterns of wave function}
\label{Section-wave-patterns}

From the QFI (\ref{Fq}) one sees that the measurement precision comes from
the sensitivity of the wave function change in response to the parameter
variation. A quicker change of the wave function driven by the parameter
variation means a larger QFI and a higher MP upper bound. The mixed linear
and nonlinear interactions in $H$ provide various variation patterns of the
wave function for the sensitivity resource.

The first pattern is the squeezing of the wave function. As illustrated in
Figs.~\ref{fig-Wave-Function}(a) and \ref{fig-Wave-Function}(b), with the
variation of $g_{2}$ in approaching to $g_{\mathrm{T}}$ the wave-function
component $\psi _{+}(x)$ tends to be narrower [Fig.~\ref{fig-Wave-Function}(a)]
while $\psi _{-}(x)$ becomes more extended [Fig.~\ref{fig-Wave-Function}(b)],
the former being position squeezing while the latter corresponding to
momentum squeezing~\cite{Ref-Squeezing}. Such squeezing effect arises from the nonlinear
coupling which renormalizes the frequency by $\varpi _{\pm }$ in the
effective harmonic potential $v_{\pm }(x)$ in Eq.~(\ref{v-potentials}).
Correspondingly $v_{+}(x)$ becomes narrower and $v_{-}(x)$ becomes broader
relatively to the original potential $v_{0}=\frac{\omega }{2}x^{2}$, as
demonstrated by Fig.~\ref{fig-v}(a). Note that the squeezing effect in $%
v_{-}(x)$ is divergently strong as $\varpi _{-}=\sqrt{1-g_{2}/g_{\mathrm{T}}}$
tends to vanish in the variation of $g_{2}$ when it approaches to $g_{\mathrm{T}}$.
Such a diverging behavior provides a resource of high
sensitivity for QM.

The second pattern is the displacement of the wave packet. As illustrated in
Fig.~\ref{fig-Wave-Function}(d), the wave packet of $\psi _{-}(x)$ is moving
quickly away from the origin ($x=0$) when $g_{2}$ increases despite that the moving
of $\psi _{-}(x)$ in the opposite direction is slower and less visible, as
driven by the potential displacement $b_{\pm }$ in the presence of a finite
linear coupling illustrated in Fig.~\ref{fig-v}(b). Indeed, the moving
acceleration of $\psi _{-}(x)$ becomes extremely large in the vicinity of
$g_{\mathrm{T}}$ as $b_{-}=g_{1}^{\prime }/(1-g_{2}/g_{\mathrm{T}})$ is
diverging. Such a diverging acceleration offers another resource of high
sensitivity for QM.

The third pattern is the transition. In the presence of the bias, as shown
in Fig.~\ref{fig-v}(c), the energy level $\varepsilon _{-}$(dotted line) can
be higher than $\varepsilon _{+}$ (dashed line) for a small $g_{2}$. An
energy level reversal can occur when $g_{2}$ is enhanced, as one finds in
Fig.~\ref{fig-v}(d), such an energy reversal induces a transition.
Corresponding the density weight is also transiting quickly from $\psi
_{+}(x)$ to $\psi _{-}(x)$, as displayed in Figs.~\ref{fig-Wave-Function}(e)
and \ref{fig-Wave-Function}(f). Such a transition supplies a third resource
of high sensitivity for QM.

Finally there is also a degeneracy-lifting pattern arising at the symmetry breaking point. In fact,
in the absence of the linear coupling, nonlinear coupling and bias field, the uncoupled and unbiased system has both spin-rotation symmetry $P_\sigma=\sigma x$ and U(1) symmetry $U(1)=e^{i\theta a^{\dagger }a}$ with an arbitrary phase $\theta$. At such a highly symmetric point the two spin energy levels at any $n$ are degenerate. Turning on the coupling e.g. the nonlinear one here, both the two symmetries are broken and the symmetry is lowered to be $P_x$ with a limited phase $\theta =\pi$. As a consequence the level degeneracy is lifted. The behavior of the wave-function variation is similar to that after the transition in Fig.~\ref{fig-v}(c), with a variation from equal density weights on the two spin components at the degenerate point to unbalanced weights at a finite $g_2$.

In the above analysis we see that the diverging origin in the squeezing
pattern and the displacement pattern lies in the $\psi _{-}(x)$ component.
For this reason, in the following we will first address the small-$\Omega $
regime, as illustrated in Figs.~\ref{fig-Wave-Function}(a)-\ref%
{fig-Wave-Function}(f), which maximizes the contribution of $\psi _{-}(x)$,
before showing the validity of these resources in the finite-$\Omega $
regime.

\section{Applications for quantum metrology}
\label{Section-QM}

\subsection{Quantum metrology by squeezing resource}
\label{Section-Squeezing}

Squeezing is a primary quantum resource that has been applied in quantum
metrology\cite{Maccone2020Squeezing,Lawrie2019Squeezing,Gietka2023PRL-Squeezing,Gietka2023PRL2-Squeezing,Candeloro2021-Squeezing}.
As described in the first pattern of the wave function in last section,
here in the nonlinear coupling of light-matter interaction the model (\ref{H-g1-g2-bias})
also manifests a squeezing effect which is divergently
strong as a sensitivity resource for QM. We can see the pure squeezing
effect at $\epsilon =0$ and $g_{1}=0$. In this case, there is no
displacement or transition and in the small-$\Omega $ regime the ground-state
wave function can be well approximated by $\psi _{\pm }(x)=c_{\pm }\varphi
_{\pm }(x)$, where $\varphi _{\pm }(x)=\xi _{\pm }^{1/4}\exp [-\frac{1}{2}\xi
_{\pm }x^{2}]/\pi ^{1/4}$ is the ground state of quantum harmonic oscillator with
renormalized frequency $\xi _{\pm }\cong \varpi _{\pm }$. In such a
situation the QFI for the parameter $g_{2}$ contains two parts
\begin{equation}
F_{Q}\left( g_{2}\right) =F_{Q}^{\xi }\left( g_{2}\right) +F_{Q}^{\rho
}\left( g_{2}\right),
\end{equation}%
respectively coming from the squeezing of the basis $\varphi _{\pm }(x)$ and
the variation of spin-component density or weight (see the derivation in
Appendix \ref{Apendix-QFI-apart}).
\begin{eqnarray}
F_{Q}^{\xi }\left( g_{2}\right)  &=&\left[ \frac{c_{+}^{2}}{8(1+\overline{g}
_{2})^{2}}+\frac{c_{-}^{2}}{8(1-\overline{g}_{2})^{2}}\right] \frac{1}{g_{
\mathrm{T}}^{2}},  \label{Fq-squeezing-hz=g1=0} \\
F_{Q}^{\rho }\left( g_{2}\right)  &=&\frac{\left( \overline{g}_{2}dw+8w_{2}
\overline{w}^{3}\right) ^{2}S_{\Omega }^{2}\omega ^{2}}{4w_{2}^{4}\left(
16S_{\Omega }^{2}+dw^{2}\omega ^{2}\right) ^{2}\overline{w}^{4}}\frac{1}{g_{
\mathrm{T}}^{2}},  \label{Fq-density-hz=g1=0}
\end{eqnarray}
where%
\begin{eqnarray}
S_{\Omega }^{2} &=&(\frac{\Omega }{2}\langle \varphi _{+}|\varphi _{-}\rangle)^2
                 =w_{2}^{1/2}\Omega ^{2}/\left( 4\overline{w}\right) , \\
c_{+}^{2} &=&\frac{\frac{1}{2}\left( \sqrt{\omega ^{2}dw^{2}+16S_{\Omega
}^{2}}-\omega dw\right) ^{2}}{16S_{\Omega }^{2}+\omega ^{2}dw^{2}-\omega dw
\sqrt{\omega ^{2}dw^{2}+16S_{\Omega }^{2}}}, \\
c_{-}^{2} &=&\frac{8S_{\Omega }^{2}}{16S_{\Omega }^{2}+\omega
^{2}dw^{2}-\omega dw\sqrt{\omega ^{2}dw^{2}+16S_{\Omega }^{2}}},
\end{eqnarray}
and we have defined $\overline{w}=(\varpi _{+}+\varpi _{-})/2$, $dw=(\varpi
_{+}-\varpi _{-})$, $w_{2}=\varpi _{+}\varpi _{-}$ and $\varpi _{\pm }=\sqrt{
1\pm \overline{g}_{2}}$.

%%%%%%%%%%%%%%%%%%%%%%%%%%%%%%%%%%%%%%%%%%%%%%%%%%%%%%%%%%%%%%%%%%%%%%%%%%%%%%%%%%%%%%%%%%%%%%%%%%
\begin{figure*}[t]
\includegraphics[width=2\columnwidth]{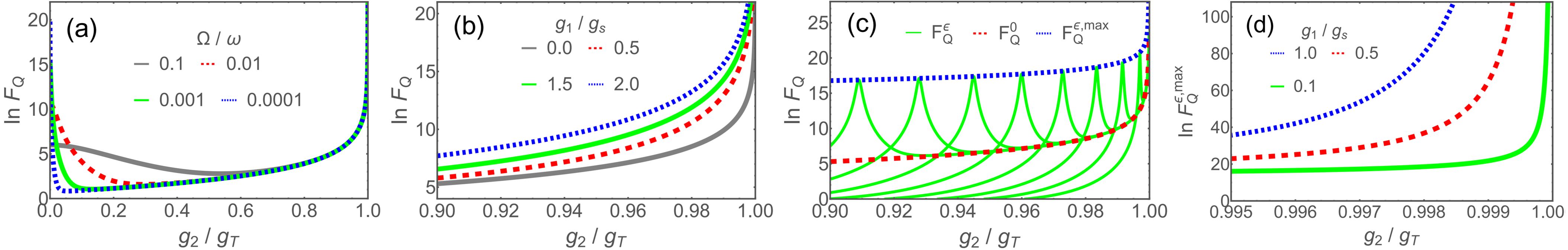}
\caption{The quantum Fisher information (QFI) $F_Q$ (plotted in natural logarithm) in the presence of different resources.
(a) $g_1=0$ and $\epsilon =0$ at different frequency ratio $\Omega/\omega=0.0001,0.001,0.01,0.1$ with squeezing resource at finite $g_2$ and degeneracy-lifting resource around $g_2=0$.
(b) $\epsilon =0$ and $\Omega=0.01\omega$ at different linear couplings $g_1/g_{\rm s}=0.0,0.5,1.5,2.0$ with squeezing and displacement resources.
(c) the QFI $F_Q^\epsilon$ by different $\epsilon/\omega=0.27 \sim 0.34$ [green (light gray) solid] in spacing $0.01$ at $g_1=0.1g_{\rm s}$ and $\Omega=0.001\omega$,  with squeezing, displacement and transition resources. The red dashed line denotes the QFI $F_Q^0$ without transition resource ($\epsilon=0$), while the blue dotted line provides the peak value  $F_Q^{\epsilon,max}$ continuously yielded by tuning $\epsilon$.
(d) $F^{\epsilon,max}_Q$ at different linear couplings $g_1/g_{\rm s}=0.1,0.5,1.0$ at $\Omega=0.01\omega$.
In all panels, $\omega=1$ is set to be the unit.}
\label{fig-QFI-g1g2hz}
\end{figure*}
%%%%%%%%%%%%%%%%%%%%%%%%%%%%%%%%%%%%%%%%%%%%%%%%%%%%%%%%%%%%%%%%%%%%%%%%%%%%%%%%%%%%%%%%%%%%%%%%%%
%%%%%%%%%%%%%%%%%%%%%%%%%%%%%%%%%%%%%%%%%%%%%%%%%%%%%%%%%%%%%%%%%%%%%%%%%%%%%%%%%%%%%%%%%%%%%%%%%%
\begin{figure*}[t]
\includegraphics[width=2\columnwidth]{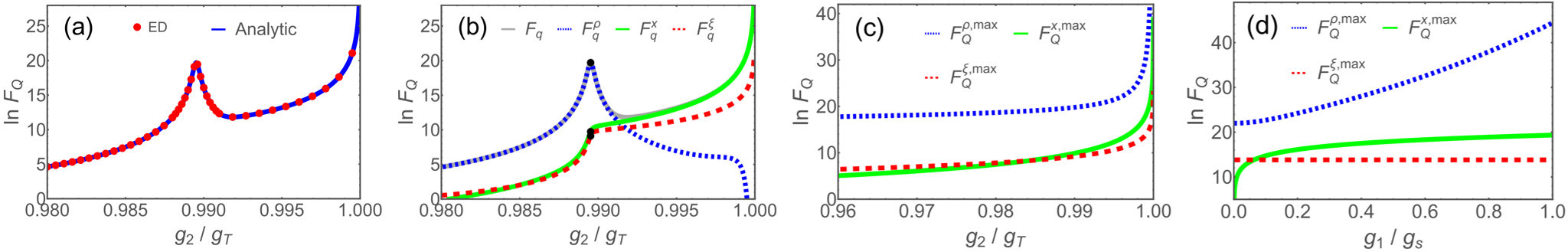}
\caption{Tracking the contributions of different sensitivity sources.
(a) Comparison of total $F_q$ for the analytic result (blue solid line) and the exact diagonalization (ED) result (red dots).
(b) Contributions of $F^\xi_Q$ (red dashed), $F^x_Q$ (green solid) and $F^\rho_Q$ (blue dotted) in total $F_Q$ (gray solid). The black dots in (b) mark the analytic $F^{\xi,max}_Q$, $F^{x, max}_Q$ and $F^{\rho,max} _Q$ at the transition point in Eqs. (\ref{Fmax}). Here $\epsilon=0.33\omega$ and $g_1=0.5g_{\rm s}$.
(c) Evolution of $F^{\xi,max}_Q$, $F^{x, max}_Q$ and $F^{\rho,max} _Q$ versus $g_2$ at fixed $g_1=0.5g_{\rm s}$.
(d) Evolution of $F^{\xi,max}_Q$, $F^{x, max}_Q$ and $F^{\rho,max} _Q$ versus $g_1$ at fixed $g_2=0.999g_{\rm T}$.
We set $\Omega =0.001\omega$ and the unit $\omega=1$ in all panels.}
\label{fig-Fq-apart}
\end{figure*}
%%%%%%%%%%%%%%%%%%%%%%%%%%%%%%%%%%%%%%%%%%%%%%%%%%%%%%%%%%%%%%%%%%%%%%%%%%%%%%%%%%%%%%%%%%%%%%%%%%

\subsubsection{Criticality of QFI}

In approaching $\overline{g}_{2}=1$ both $F_{Q}^{\xi }\left( g_{2}\right) $
and $F_{Q}^{\rho }\left( g_{2}\right) $ are diverging in a critical way%
\begin{eqnarray}
F_{Q}^{\xi }\left( \overline{g}_{2}\right)  &\sim &\frac{1}{\left( 1-
\overline{g}_{2}\right) ^{2}}\frac{1}{\omega ^{2}}, \\
F_{Q}^{\rho }\left( g_{2}\right)  &\sim &\frac{1\ }{\left( 1-\overline{g}
_{2}\right) ^{7/4}}\frac{\Omega ^{2}}{\omega ^{4}}.
\label{F-density-hz=g1=0-Kernal}
\end{eqnarray}%
The critical exponents, defined by $F_{Q}^{i}\left( g_{2}\right) \sim \left(
1-\overline{g}_{2}\right) ^{-\gamma _{i}}$, are
\begin{equation}
\gamma _{\xi }=2,\qquad \gamma _{\rho }=7/4,  \label{criti-exponent-g1=0}
\end{equation}%
respectively. As afore-mentioned, the critical behavior of $F_{Q}^{\xi }\left( \overline{g}_{2}\right)$ is the
result of the basis variation with vanishing frequency $\varpi _{-}$ which
leads to a divergently strong momentum squeezing. The diverging behavior of
$F_{Q}^{\rho }\left( g_{2}\right) $ arises from the $w_{2}^{4}$ factor,
partially canceled by $w_{2}^{1/2}$ in $S_{\Omega }^{2}$, which is the
result of the variation speed of the wave-packet overlap $S_{\Omega }=\frac{\Omega }{2}\langle \varphi _{+}|\varphi _{-}\rangle$ and
also originates from the squeezing effect. Thus, here we have a high sensitivity
resource for QM purely from squeezing.

\subsubsection{Universality of QFI}

It should be noted that $F_{Q}^{\xi }\left( \overline{g}_{2}\right) $ is
more dominant than $F_{Q}^{\rho }\left( g_{2}\right) $ in the above case. Although the magnitudes of the critical components $\gamma
_{\xi }$ in $F_{Q}^{\xi }\left( \overline{g}_{2}\right) $ and $\gamma _{\rho
}$ in $F_{Q}^{\rho }\left( g_{2}\right) $ are comparable, the contribution
of $F_{Q}^{\rho }\left( g_{2}\right) $ is diminished by the $\Omega ^{2}$
factor in the small-$\Omega $ regime. Physically, in the increase of $%
\overline{g}_{2}$, the system becomes nearly fully polarized due to the
splitting frequencies $\varpi _{\pm }$ and the weak spin flipping strength $%
\Omega $. Consequently, the final diverging behavior of the total QFI
$F_{Q}\left( g_{2}\right) $ in the vicinity of $\overline{g}_{2}=1$ is
dominated by the singular behavior of $F_{Q}^{\xi }\left( \overline{g}
_{2}\right) $ which is little affected by $\Omega $. This accounts for the
universality of $\ln F_{Q}\left( g_{2}\right) $ in different $\Omega $
values, as we see in Fig.\ref{fig-QFI-g1g2hz}(a) in the regime $\overline{g}
_{2}\in \lbrack 0.6,1]$.

It is worthy to mentioned that the criticality and universality of QFI were
noticed around the exceptional points in non-Hermitian Jaynes-Cummings
Models~\cite{Ying-Topo-JC-nonHermitian-Fisher} as another fundamental model of
light-matter interaction. Here we have the Hermitian situation. The
criticality of QFI promises a divergently high measurement precision, while
the universality of QFI guarantees a same high order of measurement
precision for different tunneling or spin flipping strengths.

\subsection{Quantum metrology by degeneracy lifting}
\label{Section-degeneracy-lifting}

In Fig.\ref{fig-QFI-g1g2hz}(a) we also see a diverging-like behavior of the
QFI in small-$\overline{g}_{2}$ regime for small $\Omega $ values. Such a
diverging-like behavior comes from $F_{Q}^{\rho }\left( g_{2}\right) $, via
the key factor $F_{Q}^{\rho }\left( g_{2}\right) \sim \left( 16\frac{%
S_{\Omega }}{\omega }+dw^{2}\frac{\omega }{S_{\Omega }}\right) ^{-2}\frac{1}{
g_{\mathrm{T}}^{2}}$, which is varying in a singular-like manner
\begin{equation}
F_{Q}^{\rho }\left( g_{2}\right) \sim \frac{1}{\overline{g}_{2}^{4}}\frac{1}{
g_{\mathrm{T}}^{2}},\text{\quad for }\Omega ^{2}\ll \overline{g}
_{2}^{2}\omega ^{2},
\end{equation}%
with respect to $\overline{g}_{2}$ and a maximum value
\begin{equation}
F_{Q}^{\rho }\left( g_{2}\right) =\frac{\omega ^{2}}{4\Omega ^{2}}\frac{1}{
g_{\mathrm{T}}^{2}},\text{\quad at \ }\overline{g}_{2}=0,
\end{equation}
diverging with respect to the small-$\Omega $ limit.

In fact, these behaviors come from the degeneracy lifting in symmetry breaking mentioned in Sec.~\ref{Section-wave-patterns}, as the energies in
the opposite spin are degenerate at $\overline{g}_{2}=0$ while the
degeneracy is lifted once a finite $\overline{g}_{2}$ is turned on. The
density weights of the two spin components become unbalanced quickly under a
weak spin flipping strength $\Omega $. This fast weight variation provides a
sensitivity resource in the small-$\overline{g}_{2}$ regime. It should be
mentioned that, although the squeezing vanishes right at $\overline{g}_{2}=0$,
there is still a finite contribution from the squeezing resource
\begin{equation}
F_{Q}^{\xi }\left( g_{2}\right) =\frac{1}{8g_{\mathrm{T}}^{2}},\text{\quad
at \ }\overline{g}_{2}=0.
\end{equation}%
This finite contribution remains due to the fact that the QFI depends on the
response to the parameter variation, rather than the parameter at one point,
while the derivative with respect to $\overline{g}_{2}$ is not vanishing
even at $\overline{g}_{2}=0$.

Since generally in QM squeezing is a more-applied resource, in the following
we shall focus more on the regime in the vicinity of $\overline{g}_{2}=1$
where the squeezing strength becomes divergently strong.

\subsection{Quantum metrology by combining squeezing and displacement resources}
\label{Section-Squeezing-Displacement}

When the linear coupling $g_{1}$ is also turned on, we get another resource from
displacement which can be combined with the squeezing resource. The linear
coupling gives rise to a displacement $b_{\pm }$ with opposite directions in
the two spin components, as in Eq.~(\ref{b-up-down}). Indeed, in the situation of $g_{1}\neq 0$ and $\epsilon =0$,
the displacement adds the contribution $F_{Q}^{x}$ to the total QFI
\begin{equation}
F_{Q}\left( g_{2}\right) =F_{Q}^{\xi }\left( g_{2}\right) +F_{Q}^{\rho
}\left( g_{2}\right) +F_{Q}^{x}\left( g_{2}\right) .
\end{equation}%
Here $F_{Q}^{\xi }$ has the same form as in (\ref{Fq-squeezing-hz=g1=0})
except for the detail in the expressions of $c_{+}^{2}$ [see Eqs. (\ref{c1c2}
)-(\ref{e-plus-minus})], while the displacement part reads (see the
derivation in Appendix \ref{Apendix-QFI-apart})
\begin{equation}
F_{Q}^{x}\left( g_{2}\right) =\left[ \frac{c_{+}^{2}}{(1+\overline{g}
_{2})^{7/2}}+\frac{c_{-}^{2}}{(1-\overline{g}_{2})^{7/2}}\right] \frac{
\overline{g}_{1}^{2}\Omega }{\omega g_{\mathrm{T}}^{2}}.
\label{Fq-x-squeez-displacement}
\end{equation}%
Note that the interplay of the linear and nonlinear couplings induces the
additional potential difference $d_{\pm }(x)$ in (\ref{v-d-g1}) which
enhances the spin polarization. In such a situation we can neglect the
$F_{Q}^{\rho }$ as there is little space left for the weight variation in almost full
polarization due to the much-weakened tunneling between the displaced wave packets.
Thus, in the leading term, the total QFI has the form
\begin{equation}
F_{Q}\left( g_{2}\right) =\left[ \frac{1}{8(1-\overline{g}_{2})^{2}}+\frac{
\overline{g}_{1}^{2}\Omega /\omega }{(1-\overline{g}_{2})^{7/2}}\right]
\frac{1}{g_{\mathrm{T}}^{2}}  \label{Fq-squeez-displacement}
\end{equation}%
as a result of the combined squeezing and displacement resources.

We see that the displacement resource yields a larger critical exponent
$\gamma _{x}$ than that of the squeezing resource,
\begin{equation}
\gamma _{\xi }=2,\qquad \gamma _{x}=7/2,
\end{equation}%
unlike the smaller one $\gamma _{\rho }=7/4$ in (\ref{criti-exponent-g1=0}).
On the other hand, the weakening factor $\Omega /\omega $ of the displacement term in (\ref{Fq-squeez-displacement}) is of lower
order than $\Omega ^{2}/\omega ^{2}$ in (\ref{F-density-hz=g1=0-Kernal}).
For these reasons $F_{Q}^{x}$ can add a contribution even larger than
$F_{Q}^{\xi }$ despite the weakening factor $\Omega /\omega $ in the small-$\Omega $ regime.
In Fig.\ref{fig-QFI-g1g2hz}(b) we see that the QFI is
indeed enhanced by the presence of the linear coupling $g_{1}$ in comparison
with the case in the absence of $g_{1}$.

\subsection{Quantum metrology by combining squeezing, displacement and transition resources}
\label{Section-Squeezing-Displacement-Transition}

\subsubsection{Most divergent QFI by a broadest combination of the
squeezing, displacement and transition resources}

Now we propose a broadest combination of the squeezing, displacement and
transition resources. The squeezing has been produced by the nonlinear coupling
$g_{2}$ and the displacement has been driven by the linear coupling $g_{1}$
as addressed in Sections~\ref{Section-Squeezing} and \ref{Section-Squeezing-Displacement}, while here the transition can be introduced
by turning on a finite bias field $\epsilon $, as analyzed in Sec.~\ref{Section-wave-patterns}. In the presence of all finite values of the linear
coupling $g_{1}$, nonlinear coupling $g_{2}$ and bias field $\epsilon $, the QFI can be also decomposed into three parts
$F_{Q}=F_{Q}^{\xi}+F_{Q}^{x}+F_{Q}^{\rho }$. Around the transition a peak of QFI emerges,
with the maximum value of the QFI also comprising of three contributions
\begin{equation}
F_{Q}^{\max }\left( g_{2}\right) =F_{Q}^{\xi ,\max }\left( g_{2}\right)
+F_{Q}^{x,\max }\left( g_{2}\right) +F_{Q}^{\rho ,\max }\left( g_{2}\right)
\end{equation}
respectively from the squeezing, displacement and transition resources.
Since the transition occurs around the level crossing which has the equal
spin-component weights, $c_{+}^{2}=c_{-}^{2}=1/2$, we obtain
\begin{eqnarray}
F_{Q}^{\xi ,\max } &=&\frac{\left( 1+\overline{g}_{2}^{2}\right) }{8(1-%
\overline{g}_{2}^{2})^{2}g_{\mathrm{T}}^{2}}, \\
F_{Q}^{x,\max } &=&\left[ \frac{1}{\left( 1-\overline{g}_{2}\right) ^{7/2}}+
\frac{1}{\left( 1+\overline{g}_{2}\right) ^{7/2}}\right] \frac{\overline{g}
_{1}^{2}\Omega }{2\omega g_{\mathrm{T}}^{2}}, \\
F_{Q}^{\rho ,\max } &=&\frac{\overline{w}\left[ w_{2}^{3}\overline{w}\omega +
\overline{g}_{1}^{2}\left( 1+\overline{g}_{2}^{2}\right) \Omega \right] ^{2}
}{4w_{2}^{17/2}\Omega ^{2}g_{\mathrm{T}}^{2}}\exp [\frac{\overline{g}
_{1}^{2}\Omega }{w_{2}^{3}\overline{w}\omega }],
\end{eqnarray}
Compared with the critical modes from the squeezing and displacement
resources
\begin{eqnarray}
F_{Q}^{\xi ,\max }\left( g_{2}\right)  &\sim &\left( 1-\overline{g}
_{2}\right) ^{-\gamma _{\xi }^{\max }}, \\
F_{Q}^{x,\max }\left( g_{2}\right)  &\sim &\left( 1-\overline{g}_{2}\right)
^{-\gamma _{x}^{\max }},
\end{eqnarray}%
with the critical exponents
\begin{equation}
\gamma _{\xi }^{\max }=2,\qquad \gamma _{x}^{\max }=7/2,
\end{equation}%
the transition resource manifests a much more divergent trend compositely
with a power divergence and an exponential divergence
\begin{equation}
F_{Q}^{\rho ,\max }\sim \left( 1-\overline{g}_{2}\right) ^{-\gamma _{\rho
}^{\max }}\exp [\frac{\overline{g}_{1}^{2}\Omega }{\overline{w}\omega \varpi
_{+}^{3}}\left( 1-\overline{g}_{2}\right) ^{-\gamma _{\rho ,\exp }^{\max }}]
\label{Fq-power-exp-divergence}
\end{equation}%
with the critical components
\begin{equation}
\gamma _{\rho }^{\max }=17/4,\qquad \gamma _{\rho ,\exp }^{\max }=3.
\end{equation}

In Fig.\ref{fig-QFI-g1g2hz}(c), the green solid lines show the evolutions of
$F_{Q}\left( g_{2}\right) $ in some continual values of the bias field $\epsilon $
under given finite values of $g_{2}$ and $g_{1}$, each having a
peak of QFI. A continuous tuning of the bias field yields the maximum QFI
$F_{Q}^{\max }$ (blue dotted line). One sees that the resource combination
with the transition achieves several orders of improvement of the QFI over
the uncombined case (red dashed line). Far more orders of improvements can
be realized by further increasing of $g_{1}$, as demonstrated in Fig.\ref%
{fig-QFI-g1g2hz}(d).

\subsubsection{Tracking the contributions of the difference resources}

The results presented in Fig.~\ref{fig-QFI-g1g2hz}(c) and \ref{fig-QFI-g1g2hz}(d) are the total QFI.
To give a better view of the resource combination, we
track the contributions of $F_{Q}^{\xi }\left( g_{2}\right) $, $F_{Q}^{\rho
}\left( g_{2}\right) $, and $F_{Q}^{x}\left( g_{2}\right) $ separately in
Fig.~\ref{fig-Fq-apart}. To confirm the analytic analysis in the above we first compare in
Fig.~\ref{fig-Fq-apart}(a) the analytic result (blue solid) and
the exact diagonalization result (ED, red dots) of $F_{Q}$, which shows a complete
agreement. The full analytic expressions for $F_{Q}^{\xi }$, $F_{Q}^{x}$, $F_{Q}^{\rho }$, and $F_{Q}$
are obtained in Appendix~\ref{Apendix-QFI-apart}, while a brief description of the ED method for
QFI~\cite{Ying-2018-arxiv,Ying2020-nonlinear-bias,Ying-gC-by-QFI-2024} is presented
in Appendix~\ref{Apendix-ED}. The separate contributions of the three
resources to the QFI are plotted in Fig.~\ref{fig-Fq-apart}(b). This
contribution tracking shows the divergent tendency of $F_{Q}^{\xi }$ (red
dashed) and $F_{Q}^{x}$ (green solid), while the peak profile in the total QFI $F_{Q}$
(gray solid) really comes from the transition resource in $%
F_{Q}^{\rho }$ (blue dotted). The black dots at the peak position mark the
analytic values of $F_{Q}^{\xi ,\max }$, $F_{Q}^{x,\max }$, and $F_{Q}^{\rho
,\max }$ which match the results of $F_{Q}^{\xi }$, $F_{Q}^{x}$, and
$F_{Q}^{\rho }$. The evolutions of $F_{Q}^{\xi ,\max }$, $F_{Q}^{x,\max }$,
and $F_{Q}^{\rho ,\max }$ are displayed in Fig.~\ref{fig-Fq-apart}(c), which
shows the differences in their divergence strengths, with achievable higher
orders of $F_{Q}^{\rho ,\max }$ over the competing $F_{Q}^{\xi ,\max }$ and
$F_{Q}^{x,\max }$. The order differences can be tuned by $g_{1}$ as shown in
Fig.~\ref{fig-Fq-apart}(d).

\subsubsection{Comparison with combination of the squeezing and transition
resources.}

In the previous sections we have added and combined the sensitivity
resources step by step from squeezing, displacement, and transition. There
is a combinition missing between the squeezing and transition resources
which may be worthy to compare. Setting $\overline{g}_{1}=0$ gives the
result of such combination of the squeezing and transition resources without
the displacement%
\begin{eqnarray}
F_{Q}^{\xi ,\max }\left( g_{2}\right)  &=&\frac{\left( 1+\overline{g}%
_{2}^{2}\right) }{8(1-\overline{g}_{2}^{2})^{2}g_{\mathrm{T}}^{2}}, \label{F-xi-g2hz} \\
F_{Q}^{\rho ,\max }\left( g_{2}\right)  &=&\frac{\overline{w}^{3}\omega ^{2}
}{4\left( 1-\overline{g}_{2}^{2}\right) ^{5/4}\Omega ^{2}}\frac{1}{\ g_{
\mathrm{T}}^{2}}.  \label{F-rho-g2hz}
\end{eqnarray}%
Despite that the critical exponent $\gamma _{\rho }^{\max }$ from the
transition resource is smaller than that from the squeezing resource, as
\begin{equation}
\gamma _{\rho }^{\max }=5/4,\qquad \gamma _{\rho ,\exp }^{\max }=2,
\end{equation}%
here in (\ref{F-xi-g2hz}) and (\ref{F-rho-g2hz}),
$F_{Q}^{\rho ,\max }\left( g_{2}\right) $ is greatly amplified by the enhancing factor
$\omega ^{2}/\Omega ^{2}$ with a strong divergence with respect to the small-$\Omega $ limit.
Still, compared with the result in (\ref{Fq-power-exp-divergence}), we see that the broadest combination with the
displacement resource provides a strongest boost for the enhancement of the
QFI.

%%%%%%%%%%%%%%%%%%%%%%%%%%%%%%%%%%%%%%%%%%%%%%%%%%%%%%%%%%%%%%%%%%%%%%%%%%%%%%%%%%%%%%%%%%%%%%%%%%
\begin{figure*}[t]
\includegraphics[width=2.00\columnwidth]{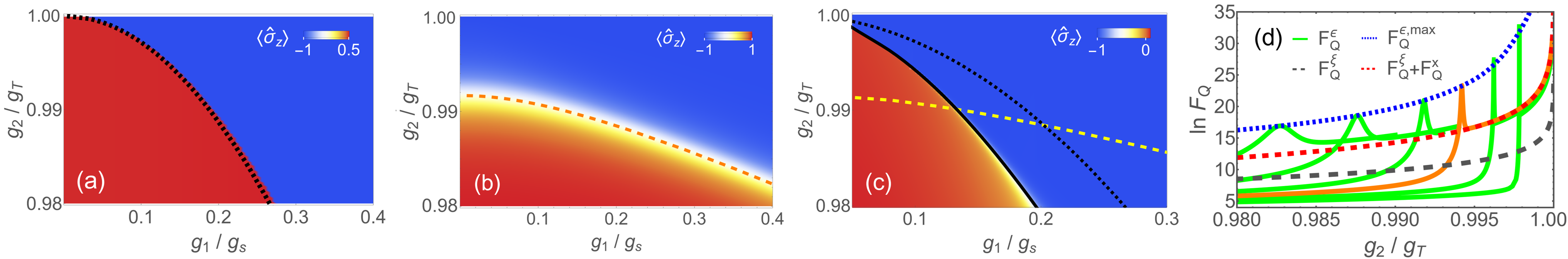}
\caption{Breaking the frequency limitation.
(a-c) Phase diagram of $\langle \widehat{\sigma} _z\rangle$ in the $g_1$-$g_2$ plane at
$\omega/\Omega=0.01$ (a),
$\Omega/\omega=0.01$ (b), and
$\omega/\Omega=1.0$ (c).
(d) Maximum QFI $F_q^{\epsilon,max}$ over $F_q^\epsilon$ (green solid) in varying $g_1$ at $\omega/\Omega=1.0$. In (d) the example of $F_q^\epsilon$ at $g_1=0.1g_{\rm s}$ (orange solid) is compared with $F_q^{x}+F_q^{\xi}$  (red dashed) and $F_q^{\xi}$ (gray long-dashed).  In all panels $\epsilon =0.33$ and $max\{\omega,\Omega\}=1$ is set to be the unit.
}
\label{fig-Finite-Omega}
\end{figure*}
%%%%%%%%%%%%%%%%%%%%%%%%%%%%%%%%%%%%%%%%%%%%%%%%%%%%%%%%%%%%%%%%%%%%%%%%%%%%%%%%%%%%%%%%%%%%%%%%%%
%%%%%%%%%%%%%%%%%%%%%%%%%%%%%%%%%%%%%%%%%%%%%%%%%%%%%%%%%%%%%%%%%%%%%%%%%%%%%%%%%%%%%%%%%%%%%%%%%%
\begin{figure*}[t]
\includegraphics[width=2.0\columnwidth]{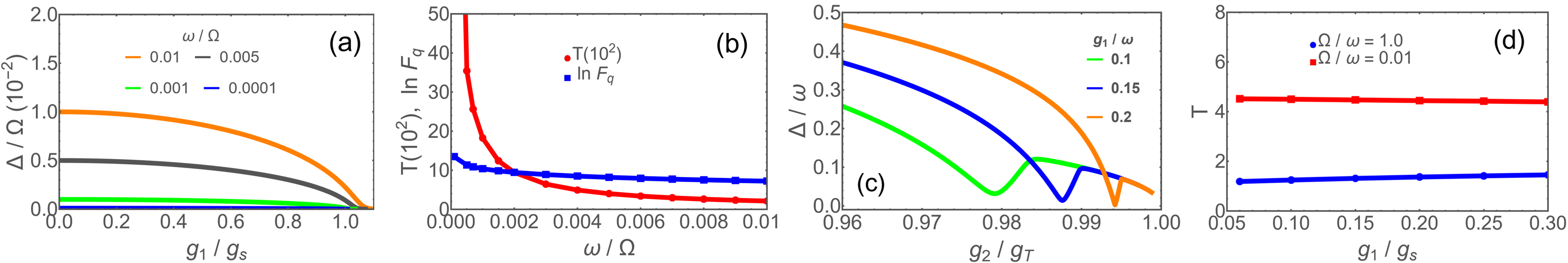}
\caption{Reduction of the preparation time of the probe state (PTPS).
(a) Gap $\Delta$ in linear interaction at low frequencies.
(b) Gap $F_Q$ (blue squares) and PTPS T in unit of $10^2$ (red dots) in linear interaction at low frequencies.
(c) Gap $\Delta$ in linear interaction in the presence of different $g_1$ at $\Omega /\omega =1$.
(d) T (in unit of $1$) versus $g_1$ at (red dots) at $\Omega /\omega =1$ (blue dots) and  $\Omega /\omega =0.01$ (red squares).
}
\label{fig-gap-time}
\end{figure*}
%%%%%%%%%%%%%%%%%%%%%%%%%%%%%%%%%%%%%%%%%%%%%%%%%%%%%%%%%%%%%%%%%%%%%%%%%%%%%%%%%%%%%%%%%%%%%%%%%%

%%%%%%%%%%%%%%%%%%%%%%%%%%%%%%%%%%%%%%%%%%%%%%%%%%%%%%%%%%%%%%%%%%%%%%%%%%%%%%%%%%%%%%%%%%%%%%%%%%
\begin{figure*}[t]
\includegraphics[width=1.8\columnwidth]{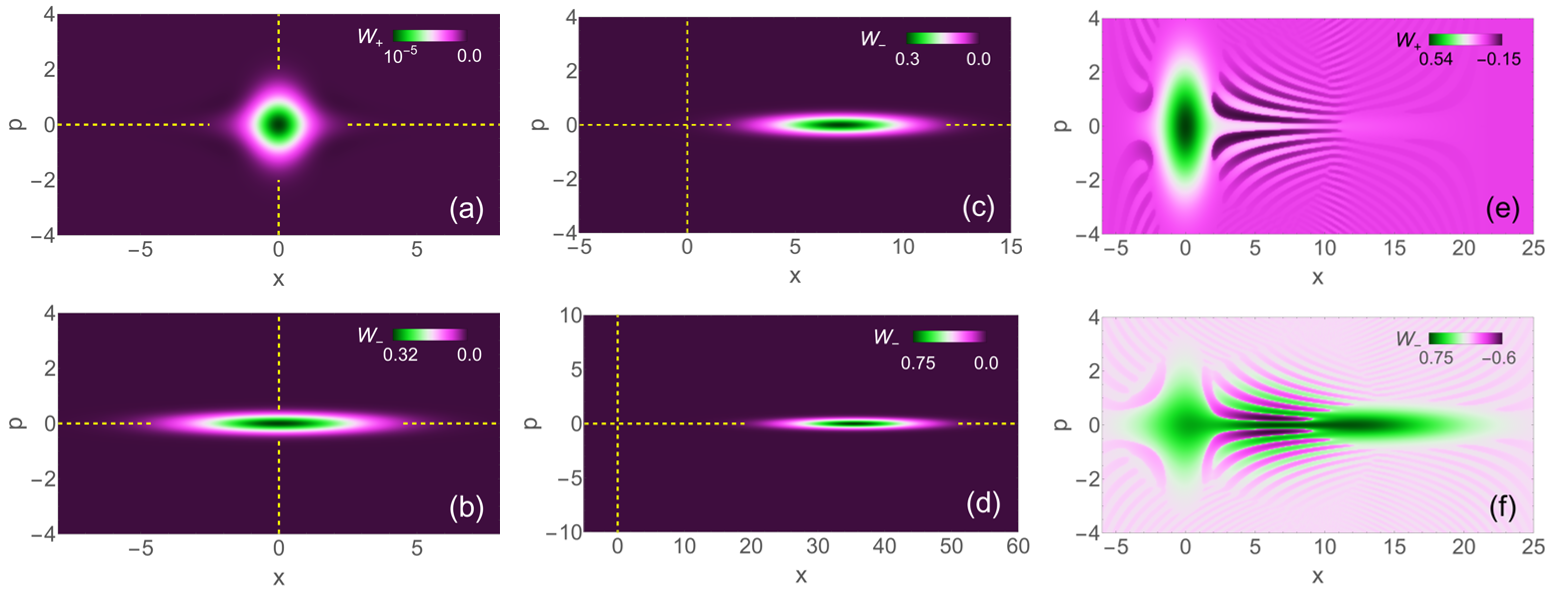}
\caption{The Wigner function $W_{\sigma}(x,p)$ in the presence of different resources.
Squeezing resource: (a,b) $g_1=0$, $\epsilon =0$, $g_2=0.99g_{\rm T}$ and $\Omega =0.01\omega$ for $\sigma=+$ (a) and $\sigma=-$ (b).
Squeezing and displacement resources: (c) $\sigma=-$, $g_1=1.0g_{\rm s}$, $\epsilon =0$, $g_2=0.99g_{\rm T}$ and $\Omega =0.01\omega$.
Squeezing, displacement and transition resources: (d) after transition, $\sigma=-$, $g_1=1.0g_{\rm s}$, $\epsilon =0.33\omega$, $g_2=0.998g_{\rm T}$ and $\Omega =1.0\omega$.
(e,f) around transition, $\sigma=+$, $g_1=0$, $\epsilon =0$, $g_2=0.9942g_{\rm T}$ and $\Omega =1.0\omega$ for $\sigma=+$ (e) and $\sigma=-$ (f).
The crossing of the dashed lines in (a-d) marks the origin.
In all panels, $\omega=1$ is set to be the unit. In (e,f) the plotted amplitude is amplified by $|W_{\sigma}|^{1/4}$.
}
\label{fig-Wigner}
\end{figure*}
%%%%%%%%%%%%%%%%%%%%%%%%%%%%%%%%%%%%%%%%%%%%%%%%%%%%%%%%%%%%%%%%%%%%%%%%%%%%%%%%%%%%%%%%%%%%%%%%%%

\section{More advantages in the quantum metrology}
\label{Section-advantages}

Besides the upgrading of upper bound of measurement precision, the protocol
of resource combination also has several other advantages including more
global parameter regime, breaking the frequency-limit restriction, and
finite preparation time of probe state (PTPS), as addressed in the following.

\subsection{Global parameter regime}

It should be mentioned that the transition resource is more flexible for
tuning. Actually the transition may be tuned to any position by the bias and
the linear coupling. Indeed, the transition can be manipulated to occur at a
wanted value of $\overline{g}_{2}$ by adding the bias $\epsilon =\epsilon
_{\max }$ under a given $\overline{g}_{2}$ or the linear coupling $\overline{
g}_{1}=\overline{g}_{1}^{\max }$ under a given $\epsilon $:
\begin{eqnarray}
\epsilon _{\max } &=&\frac{\omega }{4}(\sqrt{1+\overline{g}_{2}}-\sqrt{1-
\overline{g}_{2}})+\frac{\overline{g}_{1}^{2}\overline{g}_{2}\Omega }{
4\left( 1-\overline{g}_{2}^{2}\right) },  \label{bias-max} \\
\overline{g}_{1}^{\max } &=&\sqrt{\frac{4\left( 1-\overline{g}%
_{2}^{2}\right) }{\overline{g}_{2}\Omega }\left[ \epsilon -\frac{\omega }{4}(
\sqrt{1+\overline{g}_{2}}-\sqrt{1-\overline{g}_{2}})\right] }.
\label{g1-max}
\end{eqnarray}%
Setting $\epsilon _{\max }=0$ and $\overline{g}_{1}^{\max }=0$ will retrieve
and connect the case of the denegeracy-lifting resource discussed in Sec.~\ref{Section-degeneracy-lifting}, thus covering the entire parameter range
of $\overline{g}_{2}$.

This degree of freedom in tuning allows a more global\ validity regime and
flexibility for manipulation of QM, in contrast to the linear critical QM
in which the transition only occurs at a single coupling point.

\subsection{Breaking the frequency-limit restriction}

The favorable condition for the linear critical QM is the
low-frequency limit $\omega /\Omega \rightarrow 0$, while at finite
frequencies the QFI becomes much diminished~\cite{Ying2022-Metrology,Ying-gC-by-QFI-2024}. Such a frequency-limit restriction
is however broken in the protocol proposed in the present work. Indeed, we have the
transition resource in all frequency regimes. Actually the transition
resource is available both in the small-$\Omega $ regime and in the
low-frequency regime, as shown by the phase diagrams of $\langle \hat{\sigma }_{z}\rangle $ in
Figs.~\ref{fig-Finite-Omega}(a) and \ref{fig-Finite-Omega}(b). The transition boundary in the small-$\Omega $ regime
has been given in (\ref{bias-max}) and (\ref{g1-max}), while the transition
boundary in the low-frequency limit is provided by~\cite{Ying2020-nonlinear-bias}
\begin{equation}
\overline{g}_{1c}=\left( 1+\frac{1}{\overline{g}_{2}}\frac{\epsilon }{\Omega
}\right) \sqrt{1-\overline{g}_{2}^{2}},
\end{equation}%
as denoted by the dotted line and dashed line in Figs.~\ref{fig-Finite-Omega}(a)
and \ref{fig-Finite-Omega}(b). It is particularly worth noting that the
transition resource also remains in the finite $\Omega $ and finite $\omega $
regime, as demonstrated by the phase diagram of $\langle \hat{\sigma }_{z}\rangle $ in Fig.~\ref{fig-Finite-Omega}(c).
In fact, all resources of squeezing, displacement and transitions survive in the finite $\Omega $ and finite $\omega $ regime,
as demonstrated by the wave function in Figs.~\ref{fig-Wave-Function}(g) and ~\ref{fig-Wave-Function}(h).

The QFI in the finite $\Omega $ and finite $\omega $ regime is illustrated
in Fig.\ref{fig-Finite-Omega}(d), with the divergence behavior and resource
contributions similar to Fig.\ref{fig-QFI-g1g2hz}(c) in the small-$\Omega $
regime. Here, a bit differently for the plots, the large-$\overline{g}_{2}$
tails of $F_{Q}\left( g_{2}\right) $ in Fig.\ref{fig-Finite-Omega}(d) do
not collapse into the same line as in Fig.\ref{fig-QFI-g1g2hz}(c), due to
that we illustrate by tuning $g_{1}$ at a fixed $\epsilon $ in the former
rather than $\epsilon $ at fixed $g_{1}$ the latter. This difference can be
seen from $F_{Q}^{x}\left( g_{2}\right) $ of the displacement resource in
(\ref{Fq-x-squeez-displacement}) and (\ref{Fq-squeez-displacement})
which depends on $g_{1}$ but is little affected by $\epsilon $.

In particular, in Fig.~\ref{fig-Finite-Omega}(d) we give a comparison for the
analytic result (\ref{Fq-squeez-displacement}) of $F_{Q}^{\xi }+F_{Q}^{x}$
(red dashed) without transition at $\epsilon =0$ and the ED result of $F_{Q}$
(orange solid) with transition at the finite $\epsilon$. We see that they
agree well with each other after the transition, which not only confirms the
analytic result by ED but also indicates that the physical picture of
resource combination in the small-$\Omega $ regime also applies for the
finite-$\Omega $ regime. On the other hand, the peak position of $%
F_{Q}\left( g_{2}\right) $ by continuously varying $g_{1}$ in Fig.~\ref{fig-Finite-Omega}(d)
is marked as the black solid line in Fig.~\ref{fig-Finite-Omega}(c), which matches well with the transition boundary of
$\langle \hat{\sigma }_{z}\rangle $ there, indicating that the peak
contribution also really comes from the transition just like in the small-$\Omega $ regime.
With these confirmations, a comparing for the long-dashed
line ($F_{Q}^{\xi }$ with only squeezing resource), the red dashed line ($F_{Q}^{\xi }+F_{Q}^{x}$
by combined squeezing and displacement resources),
and the blue dotted line ($F_{Q}^{\epsilon ,\max }$ by combined all three
resources of squeezing, displacement and transition) demonstrates the
upgrading by resource combinations also in the finite-$\Omega $ regime.

More exactly speaking, in the finite $\Omega $ and finite $\omega $ regime,
the combined resources are more entangled and, unlike in the small-$\Omega $
regime, the mixed terms $F_{Q}^{\xi ,x}$, $F_{Q}^{\xi ,\rho }$, and $%
F_{Q}^{x,\rho }$ in the QFI do not completely vanish:
\begin{equation}
F_{Q}=F_{Q}^{\xi }+F_{Q}^{x}+F_{Q}^{\rho }+F_{Q}^{\xi ,x}+F_{Q}^{\xi ,\rho
}+F_{Q}^{x,\rho },
\end{equation}%
due to the emergence of anti-polaron wave packets\cite{Ying2015}.
Nevertheless, as the intra-polaron terms in $F_{Q}^{\xi }$, $F_{Q}^{x}$, and
$F_{Q}^{\rho }$ are still playing a leading role, we also have a diverging
QFI from the different sensitivity resources qualitatively similar to the
small-$\Omega $ regime, as addressed in the above for Fig \ref%
{fig-Finite-Omega}(d). We leave the expressions and corresponding discussion
in Appendix \ref{Apendix-finite-Omega}.

\subsection{Finite gap and preparation time of probe state (PTPS)}

Another detrimental problem in the linear critical QM is the
divergent PTPS. The PTPS is inversely proportional to the gap $\Delta $ in
adiabatic preparation of the probe states, so that the PTPS can be estimated
by~\cite{Garbe2020,Ying2022-Metrology}
\begin{equation}
T=\int\limits_{0}^{\overline{g}_{\max }}\Delta \left( \overline{g}\right)
^{-1}d\overline{g}
\end{equation}%
where $\overline{g}=\overline{g}_{1}$ for the linear critical QM ($\overline{g}=
\overline{g}_{2}$ for the nonlinear coupling in the present work) and
$\overline{g}_{\max }$ is the rescaled transition point where the QFI has a
peak value~\cite{Ying-gC-by-QFI-2024}. In linear coupling, the gap tends to
overall close when the frequency is lowered as indicated in Fig.~\ref{fig-gap-time}(a),
consequently the PTPS $T$ becomes diverging in the
low-frequency limit as shown by the red dots in Fig.~\ref{fig-gap-time}(b).
In contrast, in the mixed nonlinear and linear couplings discussed in the
present work, the gap (see Appendix \ref{Apendix-Wave-function})
\begin{equation}
\Delta =2\sqrt{e_{-}^{2}+S_{\Omega }^{2}}
\end{equation}%
is finite either in the small-$\Omega $ regime or in the finite-$\Omega $
regime, as the order of gap is decided by $\omega $ [see (\ref%
{gap-small-Omega}) in Appendix \ref{Apendix-Wave-function}] which is now not
necessary to be in the low-frequency limit. We illustrate the finite gap in
Fig.~\ref{fig-gap-time}(c). As a result, the PTPS $T$ is also finite, as in
Fig.~\ref{fig-gap-time}(d). Indeed, despite that the value of $\Omega $ is reduced
by two orders, the PTPS is still in the same order of single digit, in a
sharp contrast to the much larger order (unit of $10^{2}$) and diverging
behavior of the linear coupling case in Fig.~\ref{fig-gap-time}(b). Thus, our
present QM protocol also avoids the problem of diverging PTPS.

\section{Wigner function and resources exploiting from all the position, momentum,
and spin spaces}
\label{Section-Wigner}

We have seen that the combination of the squeezing, displacement and
transition resources leads to a maximization of QFI and an upgrade of
critical QM. As a more general view of point, actually the combination
exploits the sensitivity resources from all the position, momentum, and spin
spaces, which can be seen from the Wigner function~\cite{Wigner1932,WignerReview2018}
\begin{equation}
W_{\pm }\left( x,p\right) =\frac{1}{2\pi }\int\limits_{-\infty }^{\infty
}e^{ipy}\psi _{\pm }^{\ast }(x+\frac{y}{2})\psi _{\pm }(x-\frac{y}{2})dy,
\end{equation}%
where we have set $\hbar =1$. From $W_{\pm }\left( x,p\right) $ we can
visualize all the distribution information of the position, momentum, and
spin spaces. Figures~\ref{fig-Wigner}(a) and \ref{fig-Wigner}(b) show $W_{\pm
}\left( x,p\right) $ in a pure squeezing resource, we see a strong momentum
squeezing in $W_{-}\left( x,p\right) $ along the $p$-axis direction despite
that the position squeezing in $W_{+}\left( x,p\right) $ along the $x$-axis
direction is not so obvious. Figure \ref{fig-Wigner}(b) includes both the
presences of the squeezing and displacement resources, besides the momentum
squeezing in $W_{-}\left( x,p\right) $ the wave packet distribution is
shifting away from the origin (crossing point of the dotted lines). In this
displaced case the momentum squeezing is termed as phase squeezing, while
the position squeezing is called amplitude squeezing~\cite{Ref-Squeezing}. An example in the
presence of transition resource is given in Figs.~\ref{fig-Wigner}(e) and \ref{fig-Wigner}(f),
the weight distributions of the main wave packets (red broad
regions) are transferring from $W_{+}\left( x,p\right) $ to $W_{-}\left(
x,p\right) $. On the other hand, we see that, besides the large displacement
around $x=15$ in $W_{-}\left( x,p\right) $, the amplitude and phase
squeezings are still remaining. There are some fringes between and around the
main wave packets, due to the interference of the two separating
wavepackets, the visibility of which is actually amplified by plotting $|W_{\pm }\left( x,p\right) |^{1/4}$.
Here Figures \ref{fig-Wigner}(a)-\ref{fig-Wigner}(c) are illustrated for the small-$\Omega $,
while Figures \ref{fig-Wigner}(a)-\ref{fig-Wigner}(c) show the example at a finite value of $\Omega $.
We see that the combination of squeezing, displacement and
transition paves a way to exploit and combines resources for QM from all the
position, momentum, and spin spaces.

\section{Conclusions}
\label{Section-Conclusion}

We have proposed to combine various sensitivity resources to upgrade the
upper bound of measurement precision in QM. Such protocol can be realized in
light-matter-interaction system with mixed linear and nonlinear couplings
in the presence of bias field. Indeed, the system provides several quantum
resources, including squeezing, degeneracy lifting, displacement and quantum
phase transition, all with high sensitivity. We have analytically obtained
the critical components or exponential behavior of QFI which represents the
upper bound of measurement precision. We have combined and compared these
resources step by step, each combination can give a dramatic boost to the
enhancement of the QFI. Finally a broadest combination of squeezing,
displacement and quantum phase transition yields a maximized QFI with an
improvement by many orders over the widely-used squeezing resource.

In particular, in the broadest combination as afore-mentioned, the critical
components for the diverging QFI are $\gamma _{\xi }^{\max }=2$ for the
squeezing resource, $\gamma _{x}^{\max }=7/2$ for the displacement resource,
$\gamma _{\rho }^{\max }=17/4$ for the transition resource, and the
divergence from transition resource is even enhanced exponentially with an
extra critical component $\gamma _{\rho ,\exp }^{\max }=3$ inside the
exponential factor. It is surprising to see here that the divergence of
squeezing, which is the mostly exploited resource for QM, is relatively the
least divergent resource, with the smallest critical component.
Unexpectedly, the displacement resource can yield a stronger sensitivity and
the transition resource is the strongest.

Besides the upgrading of the upper bound of measurement precision, our protocol
of QM also exhibits several other advantages: (i) the parameter regime with high measurement precision is tunable
and global, while the linear critical QM is locally limited to one
critical point in our protocol; (ii) The frequency-limit restriction in the linear
critical QM is broken; (iii) Our protocol avoids the detrimental problem of
diverging PTPS which was also encountered in the linear critical QM.

Finally we have also analyzed the sensitivity resources via the Wigner function
to gain a more general view. Indeed, our protocol exploits and combines all
the resources in momentum, position and spin spaces to maximize the MP,
while simultaneously the applicable regime is much more extended and condition restriction released.

\section*{Acknowledgment}

This work was supported by the National Natural Science Foundation of China
(Grants No. 12474358, No. 11974151, and No. 12247101).

\appendix\bigskip

\section{Wave function and gap in small-$\Omega $ regime}
\label{Apendix-Wave-function}

In the small-$\Omega $ regime the ground-state wave function can be well
approximated by $\psi _{\pm }(x)=c_{\pm }\varphi _{\pm }(x)$ where
\begin{equation}
\varphi _{\pm }(x)=\xi _{\pm }^{1/4}\exp [-\frac{1}{2}\xi _{\pm }(x\pm
x_{\pm })^{2}]/\pi ^{1/4}
\end{equation}%
is the ground state of the quantum harmonic oscillator with renormalized
frequency $\xi _{\pm }$ and displacement $x_{\pm }$. On the basis of $\varphi _{\pm }(x)$
the Hamiltonian can be represented in matrix form
\begin{equation}
H=\left(
\begin{array}{cc}
\widetilde{\varepsilon }_{+} & S_{\Omega } \\
S_{\Omega } & \widetilde{\varepsilon }_{-}
\end{array}
\right)
\end{equation}
in the subspace of the lowest energies,
with the single-particle energy in the diagonal terms
\begin{eqnarray}
\widetilde{\varepsilon }_{\pm } &=&\frac{\xi _{\pm }\omega }{4}+\left( 1\pm
\overline{g}_{2}\right) \left[ 1+2\xi _{\pm }\left( x_{\pm }-b_{\pm }\right)
^{2}\right] \frac{\omega }{4\xi _{\pm }}  \notag \\
&&-\frac{\overline{g}_{1}^{2}\Omega }{4(1\pm \overline{g}_{2})}\mp \epsilon -
\frac{\omega }{2}
\end{eqnarray}%
and the tunneling or spin-flipping energy $S_{\Omega }=\frac{\Omega }{2}
\langle \varphi _{+}|\varphi _{-}\rangle $ in the non-diagonal terms,
explicitly
\begin{equation}
S_{\Omega }=\frac{(\xi _{+}\xi _{-})^{1/4}\Omega }{\sqrt{2}(\xi _{+}+\xi
_{-})^{1/2}}\exp [-\frac{(x_{+}+x_{-})^{2}\xi _{+}\xi _{-}}{2(\xi _{+}+\xi
_{-})}].  \label{S-Omega}
\end{equation}%
The eigen energy $E^{\eta }=e_{+}+\eta \sqrt{e_{-}^{2}+S_{\Omega }^{2}}$,
where $e_{\pm }=(\widetilde{\varepsilon }_{+}\pm \widetilde{\varepsilon }
_{-})/2$, has two branches labelled by $\eta =\pm $, while the ground state
takes $\eta =-$ and%
\begin{eqnarray}
c_{+} &=&B_{+}/\sqrt{B_{+}^{2}+B_{-}^{2}},\qquad c_{-}=B_{-}/\sqrt{
B_{+}^{2}+B_{-}^{2}},  \label{c1c2} \\
B_{+} &=&(e_{-}-\sqrt{e_{-}^{2}+S_{\Omega }^{2}}),\qquad B_{-}=S_{\Omega },
\end{eqnarray}
subject to normalization condition $c_{+}^{2}+c_{-}^{2}=1$.

The energy difference of the two energy branches is
\begin{equation}
\Delta =E^{+}-E^{-}=2\sqrt{e_{-}^{2}+S_{\Omega }^{2}}  \label{gap-general}
\end{equation}%
which will the gap when there is no level crossing in the lowest excited
states.

In principle $\xi _{\pm }$ and $x_{\pm }$ are variationally determined by
minimization of $E^{-}.$ Here in the small-$\Omega $ regime the basis $\varphi _{\pm }(x)$
is little affected by the weak tunneling or spin
flipping so that we can approximately set
\begin{equation}
\xi _{\pm }=\varpi _{\pm },\qquad x_{\pm }=b_{\pm },
\label{adiabatic-approximation}
\end{equation}%
which means that the basis follows the potential adiabatically. The
single-particle energy is then simplified to be%
\begin{equation}
\widetilde{\varepsilon }_{\pm }=\varepsilon _{\pm }=\frac{1}{2}\varpi _{\pm
}\omega -\frac{\overline{g}_{1}^{2}\Omega }{4(1\pm \overline{g}_{2})}\mp
\epsilon -\frac{\omega }{2}  \label{e-plus-minus}
\end{equation}%
and the gap becomes%
\begin{equation}
\Delta =\sqrt{(\omega dw+\frac{\overline{g}_{1}^{2}\overline{g}_{2}\Omega }{
1-\overline{g}_{2}^{2}}-\epsilon )^{2}+\frac{w_{2}^{1/2}\Omega ^{2}/(16
\overline{w})}{\exp (\frac{2\overline{g}_{1}^{2}\Omega }{w_{2}^{3}\sqrt{2
\overline{w}}\omega })}}  \label{gap-small-Omega}
\end{equation}
where $\overline{w}=(\varpi _{+}+\varpi _{-})/2$, $dw=(\varpi _{+}-\varpi
_{-})$, $w_{2}=\varpi _{+}\varpi _{-}$ and $\varpi _{\pm }=\sqrt{1\pm
\overline{g}_{2}}$.

\section{QFI in small-$\Omega $ regime}
\label{Apendix-QFI-apart}

From Section \ref{Section-QFI} we have the simplified QFI for a real wave
function~\cite{Ying-gC-by-QFI-2024}
\begin{equation}
F_{Q}=4\langle \psi ^{\prime }\left( \lambda \right) |\psi ^{\prime }\left(
\lambda \right) \rangle ^{2}.
\end{equation}%
Since the wave function $\psi _{\pm }(x)=c_{\pm }\varphi _{\pm }(x)$
introduced in Appendix \ref{Apendix-Wave-function} depends on the frequency
renormalization $\xi _{\pm }$, the displacement $x_{\pm }$ and the spin
component weight $c_{\pm },$ the derivative with respect to the parameter
$\lambda =g_{2}$ includes three parts
\begin{equation}
\psi _{\pm }^{\prime }\left( g_{2}\right) =c_{\pm }\frac{d\varphi _{\pm }}{
d\xi _{\pm }}\frac{d\xi _{\pm }}{dg_{2}}+c_{\pm }\frac{d\varphi _{\pm }}{
dx_{\pm }}\frac{dx_{\pm }}{dg_{2}}+\frac{dc_{\pm }}{dg_{2}}\varphi _{\pm }.
\end{equation}%
Correspondingly the QFI contains three contributions%
\begin{equation}
F_{Q}=F_{Q}^{\xi }+F_{Q}^{x}+F_{Q}^{\rho }
\end{equation}%
wtih%
\begin{eqnarray}
F_{Q}^{\xi } &=&4\sum\limits_{\sigma =\pm }c_{\sigma }^{2}\langle \frac{
d\varphi _{\sigma }}{d\xi _{\sigma }}|\frac{d\varphi _{\sigma }}{d\xi
_{\sigma }}\rangle \left( \frac{d\xi _{\sigma }}{dg_{2}}\right) ^{2}, \\
F_{Q}^{x} &=&4\sum\limits_{\sigma =\pm }c_{\sigma }^{2}\langle \frac{%
d\varphi _{\sigma }}{dx_{\sigma }}|\frac{d\varphi _{\sigma }}{dx_{\sigma }}
\rangle \left( \frac{dx_{\sigma }}{dg_{2}}\right) ^{2}, \\
F_{Q}^{\rho } &=&4\sum\limits_{\sigma =\pm }\left( \frac{dc_{\sigma }}{dg_{2}
}\right) ^{2}=4\frac{B_{+}^{\prime }B_{-}-B_{+}B_{-}^{\prime }}{\left(
B_{+}^{2}+B_{-}^{2}\right) ^{2}},
\end{eqnarray}%
coming from the squeezing, displacement and spin weight variation
respectively. Note here that the mixed terms are vanishing
\begin{eqnarray}
F_{Q}^{\xi ,x} &=&8\sum\limits_{\sigma =\pm }c_{\sigma }^{2}\langle \frac{
d\varphi _{\sigma }}{d\xi _{\sigma }}|\frac{d\varphi _{\sigma }}{dx_{\sigma }
}\rangle \frac{d\xi _{\sigma }}{dg_{2}}\frac{dx_{\sigma }}{dg_{2}}=0, \\
F_{Q}^{\xi ,\rho } &=&8\sum\limits_{\sigma =\pm }c_{\sigma }\langle \frac{
d\varphi _{\sigma }}{d\xi _{\sigma }}|\varphi _{\sigma }\rangle \frac{d\xi
_{\sigma }}{dg_{2}}\frac{dc_{\sigma }}{dg_{2}}=0, \\
F_{Q}^{x,\rho } &=&8\sum\limits_{\sigma =\pm }c_{\sigma }\langle \frac{
d\varphi _{\sigma }}{dx_{\sigma }}|\varphi _{\sigma }\rangle \frac{
dx_{\sigma }}{dg_{2}}\frac{dc_{\sigma }}{dg_{2}}=0,
\end{eqnarray}%
as the normalization condition and $\frac{d\varphi _{\sigma }(x)}{d\xi
_{\sigma }}\frac{d\varphi _{\sigma }(x)}{dx_{\sigma }}$ as a product of even
and odd functions of $(x\pm x_{\pm })$ lead to vanishing expectations
$\langle \frac{d\varphi _{\sigma }}{d\xi _{\sigma }}|\varphi _{\sigma
}\rangle =\langle \frac{d\varphi _{\sigma }}{dx_{\sigma }}|\varphi _{\sigma
}(x)\rangle =\langle \frac{d\varphi _{\sigma }}{d\xi _{\sigma }}|\frac{
d\varphi _{\sigma }}{dx_{\sigma }}\rangle =0.$

In terms of small-$\Omega $ conditions (\ref{adiabatic-approximation}),
\begin{eqnarray}
\langle \frac{d\varphi _{\pm }}{d\xi _{\pm }}|\frac{d\varphi _{\pm }}{d\xi
_{\pm }}\rangle &=&\frac{1}{8\xi _{\pm }^{2}}=\frac{1}{8(1\pm \overline{g}
_{2})}, \\
\langle \frac{d\varphi _{\pm }}{dx_{\pm }}|\frac{d\varphi _{\pm }}{dx_{\pm }}
\rangle &=&\frac{\xi _{\pm }}{2}=\frac{\sqrt{(1\pm \overline{g}_{2})}}{2}, \\
\frac{d\xi _{\pm }}{dg_{2}} &=&\pm \frac{1}{2\sqrt{(1\pm \overline{g}_{2})}}
\frac{1}{g_{\mathrm{T}}}, \\
\frac{dx_{\pm }}{dg_{2}} &=&\mp \frac{\overline{g}_{1}\sqrt{\Omega /\omega }
}{\sqrt{2}(1\pm \overline{g}_{2})^{2}}\frac{1}{g_{\mathrm{T}}},
\end{eqnarray}
we find
\begin{eqnarray}
F_{Q}^{\xi } &=&\left[ \frac{c_{+}^{2}}{8(1+\overline{g}_{2})^{2}}+\frac{
c_{-}^{2}}{8(1-\overline{g}_{2})^{2}}\right] \frac{1}{g_{\mathrm{T}}^{2}}, \\
F_{Q}^{x} &=&\left[ \frac{c_{+}^{2}}{(1+\overline{g}_{2})^{7/2}}+\frac{
c_{-}^{2}}{(1-\overline{g}_{2})^{7/2}}\right] \frac{\overline{g}
_{1}^{2}\Omega }{\omega g_{\mathrm{T}}^{2}}, \\
F_{Q}^{\rho } &=&4\frac{B_{+}^{\prime }B_{-}-B_{+}B_{-}^{\prime }}{\left(
B_{+}^{2}+B_{-}^{2}\right) ^{2}},
\end{eqnarray}
where
\begin{eqnarray}
B_{+}^{\prime } &=&\frac{dB_{+}}{dg_{2}}=\left( e_{-}^{\prime }-\frac{
e_{-}e_{-}^{\prime }+S_{\Omega }S_{\Omega }^{\prime }}{\sqrt{
e_{-}^{2}+S_{\Omega }^{2}}}\right) \frac{1}{g_{\mathrm{T}}^{2}}, \\
B_{-}^{\prime } &=&\frac{dB_{-}}{dg_{2}}=S_{\Omega }^{\prime },
\end{eqnarray}
and%
\begin{eqnarray}
S_{\Omega } &=&\frac{w_{2}^{1/4}\Omega }{\sqrt{4\overline{w}}}\exp \left( -
\frac{\overline{g}_{1}^{2}\Omega }{w_{2}^{3}\sqrt{2\overline{w}}\omega }
\right) , \\
S_{\Omega }^{\prime } &=&-\left\{ \frac{\overline{g}_{2}}{8w_{2}^{2}%
\overline{w}^{2}}+\frac{12\overline{g}_{2}\overline{w}-\left( 7\overline{g}
_{2}^{2}-1\right) dw}{8w_{2}^{6}\overline{w}^{2}\omega /[\overline{g}%
_{1}^{2}\Omega ]}\right\} \frac{S_{\Omega }}{g_{\mathrm{T}}},  \label{dSdg2}
\\
e_{-} &=&\frac{dw}{4}\omega +\frac{\overline{g}_{1}^{2}\overline{g}_{2}}{
4w_{2}^{2}}\Omega -\epsilon , \\
e_{-}^{\prime } &=&\frac{1}{4}\left[ \frac{\overline{w}}{w_{2}}\omega +\frac{
\overline{g}_{1}^{2}\left( 1+\overline{g}_{2}^{2}\right) \Omega }{w_{2}^{4}}
\right] \frac{1}{g_{\mathrm{T}}^{2}}.
\end{eqnarray}%
Here $c_{\pm }$ is given in (\ref{c1c2}) with substitution of (\ref{adiabatic-approximation})
and we have defined $\overline{w}=(\varpi
_{+}+\varpi _{-})/2$, $dw=(\varpi _{+}-\varpi _{-})$, $w_{2}=\varpi
_{+}\varpi _{-}$ and $\varpi _{\pm }=\sqrt{1\pm \overline{g}_{2}}$.

\section{QFI in finite-$\Omega $ regime}
\label{Apendix-finite-Omega}

In the finite-$\Omega $ regime, the tunneling or spin-flipping becomes
strong so that the wave function is composed of more than one wave packets ($n_{p}\geqslant 1$)
\begin{equation}
\psi _{\pm }(x)=\sum\limits_{a=1}^{n_{p}}c_{\pm }^{a}\varphi _{\pm }^{a}(x)
\end{equation}%
where
\begin{equation}
\varphi _{\pm }^{a}(x)=\left( \xi _{\pm }^{a}\right) ^{1/4}\exp [-\frac{1}{2}
\xi _{\pm }^{a}(x\pm x_{\pm }^{a})^{2}]/\pi ^{1/4}.
\end{equation}%
The wave packets include polarons with the position $\mp x_{\pm }^{a}$
around the potential bottom $\mp b_{\pm }^{a}$ of the same spin component
and anti-polarons with the position $\mp x_{\pm }^{a}$ around the potential
bottom $\pm b_{\mp }^{a}$ of the opposite spin component. A simplest
description without loss of accuracy can be $n_{p}=2$, representing one
polaron and one anti-polaron in each spin component\cite{Ying2015}.

The variation of the wave function now includes all polarons
\begin{equation}
\frac{d\psi _{\pm }}{dg_{2}}=\sum\limits_{a}^{n_{p}}\left( c_{\pm }^{a}\frac{
d\varphi _{\pm }^{a}}{d\xi _{\pm }^{a}}\frac{d\xi _{\pm }^{a}}{dg_{2}}%
+c_{\pm }^{a}\frac{d\varphi _{\pm }^{a}}{dx_{\pm }^{a}}\frac{dx_{\pm }^{a}}{
dg_{2}}+\frac{dc_{\pm }^{a}}{dg_{2}}\varphi _{\pm }^{a}\right) .
\end{equation}%
Unlike in the small-$\Omega $ regime, the mixed terms $F_{Q}^{\xi ,x}$, $F_{Q}^{\xi
,\rho }$, and $F_{Q}^{x,\rho }$ should be picked up in the QFI
\begin{equation}
F_{Q}=F_{Q}^{\xi }+F_{Q}^{x}+F_{Q}^{\rho }+F_{Q}^{\xi ,x}+F_{Q}^{\xi ,\rho
}+F_{Q}^{x,\rho }.
\end{equation}
The pure-resource terms now include intra-polaron terms ($a=a^{\prime }$)
and inter-polaron terms ($a\neq a^{\prime }$)
\begin{eqnarray}
F_{Q}^{\xi } &=&\sum\limits_{a}^{n_{p}}\sum\limits_{a^{\prime
}}^{n_{p}}\sum\limits_{\sigma =\pm }c_{\sigma }^{a}c_{\sigma }^{a^{\prime
}}\langle \frac{d\varphi _{\sigma }^{a}}{d\xi _{\sigma }^{a}}|\frac{d\varphi
_{\sigma }^{a^{\prime }}}{d\xi _{\sigma }^{a^{\prime }}}\rangle \frac{d\xi
_{\sigma }^{a}}{dg_{2}}\frac{d\xi _{\sigma }^{a^{\prime }}}{dg_{2}}, \\
F_{Q}^{x} &=&\sum\limits_{a}^{n_{p}}\sum\limits_{a^{\prime
}}^{n_{p}}\sum\limits_{\sigma =\pm }c_{\sigma }^{a}c_{\pm }^{a^{\prime
}}\langle \frac{d\varphi _{\sigma }^{a}}{dx_{\sigma }^{a}}|\frac{d\varphi
_{\sigma }^{a^{\prime }}}{dx_{\sigma }^{a^{\prime }}}\rangle \frac{
dx_{\sigma }^{a}}{dg_{2}}\frac{dx_{\sigma }^{a^{\prime }}}{dg_{2}}, \\
F_{Q}^{\rho } &=&\sum\limits_{a}^{n_{p}}\sum\limits_{a^{\prime
}}^{n_{p}}\sum\limits_{\sigma =\pm }\langle \varphi _{\sigma }^{a}|\varphi
_{\sigma }^{a^{\prime }}\rangle \frac{dc_{\sigma }^{a}}{dg_{2}}\frac{
dc_{\sigma }^{a^{\prime }}}{dg_{2}},
\end{eqnarray}%
while the mixed terms are not vanishing due to the inter-polaron terms
\begin{eqnarray}
F_{Q}^{\xi ,x} &=&\sum\limits_{a}^{n_{p}}\sum\limits_{a^{\prime }\neq
a}^{n_{p}}\sum\limits_{\sigma =\pm }\left( c_{\sigma }^{a}c_{\sigma
}^{a^{\prime }}\langle \frac{d\varphi _{\sigma }^{a}}{d\xi _{\sigma }^{a}}|
\frac{d\varphi _{\sigma }^{a^{\prime }}}{dx_{\sigma }^{a^{\prime }}}\rangle
\frac{d\xi _{\sigma }^{a}}{dg_{2}}\frac{dx_{\sigma }^{a^{\prime }}}{dg_{2}}
\right.   \notag \\
&&\left. +c_{\sigma }^{a}c_{\sigma }^{a^{\prime }}\langle \frac{d\varphi
_{\sigma }^{a}}{dx_{\sigma }^{a}}|\frac{d\varphi _{\sigma }^{a^{\prime }}}{
d\xi _{\sigma ^{\prime }}^{a^{\prime }}}\rangle \frac{dx_{\sigma }^{a}}{
dg_{2}}\frac{d\xi _{\sigma }^{a^{\prime }}}{dg_{2}}\right) , \\
F_{Q}^{\xi ,\rho } &=&\sum\limits_{a}^{n_{p}}\sum\limits_{a^{\prime }\neq
a}^{n_{p}}\sum\limits_{\sigma =\pm }\left( c_{\sigma }^{a}\frac{d\xi
_{\sigma }^{a}}{dg_{2}}\langle \frac{d\varphi _{\sigma }^{a}}{d\xi _{\sigma
}^{a}}|\varphi _{\sigma }^{a^{\prime }}\rangle \frac{dc_{\sigma }^{a^{\prime
}}}{dg_{2}}\right.   \notag \\
&&\left. +\frac{dc_{\sigma }^{a}}{dg_{2}}\langle \varphi _{\sigma }^{a}|
\frac{d\varphi _{\sigma }^{a^{\prime }}}{d\xi _{\sigma }^{a^{\prime }}}%
\rangle c_{\sigma }^{a^{\prime }}\frac{d\xi _{\sigma }^{a^{\prime }}}{dg_{2}}
\right) , \\
F_{Q}^{x,\rho } &=&\sum\limits_{a}^{n_{p}}\sum\limits_{a^{\prime }\neq
a}^{n_{p}}\sum\limits_{\sigma =\pm }\left( c_{\sigma }^{a}\frac{dx_{\sigma
}^{a}}{dg_{2}}\langle \frac{d\varphi _{\sigma }^{a}}{dx_{\sigma }}|\varphi
_{\sigma }^{a^{\prime }}\rangle \frac{dc_{\sigma }^{a^{\prime }}}{dg_{2}}
\right.   \notag \\
&&\left. +\frac{dc_{\sigma }^{a}}{dg_{2}}\langle \varphi _{\sigma }^{a}|
\frac{d\varphi _{\sigma }^{a^{\prime }}}{dx_{\sigma }^{a^{\prime }}}\rangle
c_{\sigma }^{a^{\prime }}\frac{dx_{\sigma }^{a^{\prime }}}{dg_{2}}\right) .
\end{eqnarray}
Nevertheless, the intra-polaron terms in $F_{Q}^{\xi ,x}$, $F_{Q}^{\xi ,\rho
}$, and $F_{Q}^{x,\rho }$are still vanishing due to
\begin{eqnarray}
\langle \frac{d\varphi _{\sigma }^{a}}{d\xi _{\sigma }^{a}}|\frac{d\varphi
_{\sigma }^{a}}{dx_{\sigma }^{a}}\rangle  &=&\langle \frac{d\varphi _{\sigma
}^{a}}{d\xi _{\sigma }^{a}}|\varphi _{\sigma }^{a}\rangle =\langle \frac{
d\varphi _{\sigma }^{a}}{dx_{\sigma }}|\varphi _{\sigma }^{a}\rangle =0, \\
\langle \frac{d\varphi _{\sigma }^{a}}{dx_{\sigma }^{a}}|\frac{d\varphi
_{\sigma }^{a}}{d\xi _{\sigma }^{a}}\rangle  &=&\langle \varphi _{\sigma
}^{a}|\frac{d\varphi _{\sigma }^{a}}{d\xi _{\sigma }^{a}}\rangle =\langle
\varphi _{\sigma }^{a}|\frac{d\varphi _{\sigma }^{a}}{dx_{\sigma }^{a}}
\rangle =0.
\end{eqnarray}
As the inter-polaron overlaps in separate wave packets are relatively
smaller, the intra-polaron terms are still playing a leading role. Note that
$F_{Q}^{\xi ,x}$, $F_{Q}^{\xi ,\rho }$, and $F_{Q}^{x,\rho }$ only contain
inter-polaron terms, while $F_{Q}^{\xi }$, $F_{Q}^{x}$, and $F_{Q}^{\rho }$
are composed of larger intra-polaron terms and smaller inter-polaron terms. Furthermore, in
a finite displacement the anti-polaron has much smaller weight due to the
higher potential in a displacement direction opposite to the potential~\cite{Ying2015,Ying2020-nonlinear-bias}. As a result, the polaron
contribution described in Appendix \ref{Apendix-QFI-apart} is still playing
a leading role. Thus, in the finite-$\Omega $ regime, we also have a
diverging QFI from the different sensitivity resources qualitatively similar
to the small-$\Omega $ regime, as one sees in Fig \ref{fig-Finite-Omega}(d).

\section{QFI by exact diagonalization}
\label{Apendix-ED}

In the exact diagonalization the wave function is expanded in the Fock space
\begin{equation}
\left\vert \psi \right\rangle =\left\vert \psi _{+}\right\rangle +\left\vert
\psi _{-}\right\rangle ,\qquad \left\vert \psi _{\pm }\right\rangle
=\sum\limits_{n=0}^{\infty }c_{n}^{\pm }\left\vert n\right\rangle
\end{equation}
where $n$ is the photon number and $\pm $ labels the spin components. The
Hamiltonian can be rewritten in a matrix form on the basis $\left\vert \pm
,n\right\rangle $, the matrix of the mixed linear and nonlinear couplings
with the bias field is provided in Ref.\cite{Ying2020-nonlinear-bias}. In
principle, the basis sums over all photon number, while a basis cutoff is
applied in practice within the convergence in required precision. The derivative of the
wave function with respect to the parameter $\lambda $ is then determined by
the variations of the basis weights $c_{n}^{\pm }$
\begin{equation}
\left\vert \psi _{\pm }^{\prime }\right\rangle =\sum\limits_{n=0}^{\infty }
\frac{dc_{n}^{\pm }}{d\lambda }\left\vert n\right\rangle .
\end{equation}
Then the QFI for a pure state is available by
\begin{eqnarray}
F_{Q}\left( \lambda \right)  &=&4\left[ \langle \psi ^{\prime }\left(
\lambda \right) |\psi ^{\prime }\left( \lambda \right) \rangle -\left\vert
\langle \psi ^{\prime }\left( \lambda \right) |\psi \left( \lambda \right)
\rangle \right\vert ^{2}\right]   \notag \\
&=&4\left( \sum\limits_{n=0}^{\infty }\left\vert \frac{dc_{n}^{\pm }}{
d\lambda }\right\vert ^{2}-\left\vert \sum\limits_{n=0}^{\infty }\frac{
dc_{n}^{\pm \ast }}{d\lambda }c_{n}^{\pm }\right\vert ^{2}\right) \label{Fq-ED-Cn}
\end{eqnarray}%
As mentioned in Sec.~\ref{Section-QFI}, the $\left\vert \langle \psi ^{\prime }\left(
\lambda \right) |\psi \left( \lambda \right) \rangle \right\vert $ term in $
F_{Q}\left( \lambda \right) $ vanishes for a real wave function~\cite{Ying-gC-by-QFI-2024}, which is exact and also can be checked numerically by
the exact diagonalization via the second term in (\ref{Fq-ED-Cn}) within the required numerical precision.

\end{document}